# Near-field thermal emission from metasurfaces constructed of SiC ellipsoidal particles

Lindsay P. Walter[†,*], Joseph C. McKay[†], Bart Raeymaekers[‡], and Mathieu Francoeur[‡‡,*]

[†]Radiative Energy Transfer Lab, Department of Mechanical Engineering, University of Utah, Salt Lake City, UT 84112, USA

[‡]Manufacturing and Tribology Laboratory, Department of Mechanical Engineering, Virginia Tech, Blacksburg, VA 24061, USA

[‡‡]Department of Mechanical Engineering, McGill University, Montréal, QC H3A 0C3, Canada

[*]Corresponding authors: L.Walter@utah.edu, mathieu.francoeur@mcgill.ca

**ABSTRACT:**

We model near-field thermal emission from metasurfaces structured as two-dimensional arrays of ellipsoidal SiC particles. The modeling approach is developed from fluctuational electrodynamics and is applicable to systems of ellipsoidal particles within the dipole limit. In all simulations, the radial lengths of particles are restricted to the range of 10 to 100 nm, and interparticle spacing is constrained to at least three times the particle characteristic length. The orientation and dimensions of constituent ellipsoidal particles are varied to tune localized surface phonon resonances and control the near-field energy density above metasurfaces. Results show that particle orientation can be used to regulate the relative magnitude of resonances in the energy density and particle dimensions may be changed to adjust the frequency of these resonances within the Reststrahlen band. Metasurfaces constructed from particles with randomized dimensions display comparatively broadband thermal emission rather than the three distinct resonances seen in metasurfaces made with ellipsoidal particles of equivalent dimensions. When the interparticle spacing in a metasurface exceeds about three times the particle characteristic length, the spectral energy density above the metasurface is dominated by individual particle self-interaction and can be approximated as a linear combination of single-particle spectra. When interparticle spacing is at the lower limit of three times the characteristic length, however, multiparticle interaction effects increase, and the spectral energy density above a metasurface deviates from that of single particles. This work provides guidance for designing all-dielectric, particle-based metasurfaces with desired near-field thermal emission spectra, such as thermal switches.

## INTRODUCTION

Metamaterials are engineered materials composed of structures much smaller than the operational electromagnetic wavelength and display properties unique from those of their constituent materials [1]. In thermal applications, metamaterials have been designed to tune the spectrum [2–4] and direction [5–8] of thermal emission for desired functionalities. For example, thermal metamaterials have been designed for both fixed and dynamic emissivity regulation in daytime radiative cooling applications [9–13], to increase efficiency of thermophotovoltaic devices through selective-wavelength emitter design [14–17], as a thermal lens for focused heating over a delimited spot [18], and for thermal camouflage materials [19–22]. To achieve such a broad range in functionalities, researchers have worked to develop efficient methods and strategies for metamaterial design [23–25].

For full characterization and informed design of metamaterials across length scales, it is important to resolve thermal radiation in the near field [26]. Overwhelmingly, researchers have focused on modeling near-field thermal radiation from metamaterials structured as one- and two-dimensional gratings composed of plasmonic materials, such as metals and doped semiconductors [27–35]. However, the design space of grating structures is somewhat limited (i.e., tuning thermal radiation is restricted to material modification and adjustments of grating size parameters). In order to realize a diverse set of metamaterial functionalities, alternative metamaterial designs are required.

One metamaterial structure that has shown promising thermal behavior is the particle-based structure [36–38]. Researchers have found that particle-based metamaterials can support collective lattice resonances that affect thermal emission [39] and can display enhanced heat flux in the near field due to excitation of localized surface modes [40,41]. Localized surface modes do not exist in bulk materials and are affected by particle geometry and interparticle interactions. In all-dielectric, particle-based metamaterials, localized surface phonons (LSPhs) can be the dominant mode of near-field thermal emission [42], thereby providing opportunity to tune thermal emission through geometric modification of the metamaterial structure. Such particle-based metamaterials could be manufactured via ultrasound directed self-assembly which allows three-dimensional control over the spatial organization of the particles dispersed in a macroscale volume [43–45].

To handle the scale of particle-based metamaterials, near-field thermal radiation models must support large numbers of particles. Since fluctuational electrodynamics models of near-field thermal radiation are often computationally intractable for metamaterials made of many unique, complex-shaped particles, simplified dipole approximations have been developed to reduce computational costs in particle-based designs [46–48]. Researchers have employed these dipole approximations to model near-field thermal radiation in systems of many particles arranged in random clusters [49], in ordered particle arrays [50–53], and in fractal formations [54]. Thus far, however, models between large groups of particles have been restricted to spherical dipoles. Some researchers have modeled near-field radiative heat transfer between nonspherical dipoles, such as spheroids [55–57], but these analyses have been limited to systems of three or fewer particles. To innovate new particle-based metamaterial designs, dipole models must accurately reflect the particle distributions and geometric irregularities in real, manufactured metamaterials. Development of nonspherical dipole models for near-field thermal radiation between large numbers of particles would help to realize this outcome.

In this article, we address this knowledge gap by presenting a model of near-field thermal emission above dielectric-based metasurfaces constructed of ellipsoidal particles. By focusing on ellipsoidal particles, we can control the geometry of particles along three separate axes, thereby broadening the scope of what types of particles may be modeled, from needle-like structures to asymmetric flat disks and beyond. Here, metasurfaces are composed of a single layer of ellipsoidal SiC particles with particle radii between 10 and 100 nm. Interparticle spacing is at least three times the particle characteristic length, a regime in which the accuracy of dipole approximations is deemed acceptable [48]. We vary the particle dimensions and orientation and analyze the resulting near-field thermal emission as characterized by the spectral near-field energy density. Near-field energy density is modeled using a variation of the discrete system Green's function method [58,59] where ellipsoidal geometry is accounted for in the self-interaction term of the free-space Green's function. We find that proper choice of geometric parameters (e.g., dimensions, orientation) of the constituent ellipsoidal particles allows for spectral control and tunability of near-field thermal emission from a metasurface. In particular, the particle dimensions can be used to control the location and bandwidth of localized LSPh resonances that contribute to thermal emission within the Reststrahlen band.

## METHODS

### Dipole approximation.

We implement an extension of the many-body theory of near-field radiative heat transfer [47,50,60,61] derived within the framework of the discrete system Green's function method [58,59] to calculate the energy density above metasurfaces composed of ellipsoidal particles in the dipole limit. For application of dipole approximations, we require two constraints: firstly, the characteristic length of the particles must be smaller than the thermal wavelength defined by Wien's law (i.e., $L_{\text{ch}} \ll \lambda_T = 2898/T$ μm·K), and, secondly, the center-of-mass separation distance between particles must be at least three times the characteristic length of the particles (i.e., $d \gtrsim 3L_{\text{ch}}$) [48]. For ellipsoidal particles, the characteristic length is defined as the maximum of the three ellipsoid semiaxes, $L_{\text{ch}} = \max(\{a, b, c\})$. The three ellipsoid semiaxes $a$, $b$, and $c$ are given as aligned with the $x$-, $y$-, and $z$-axes of the local Cartesian coordinate system, such that the equation for the ellipsoid quadratic surface may be written as $\frac{x^2}{a^2} + \frac{y^2}{b^2} + \frac{z^2}{c^2} = 1$. In this work, the semiaxes are defined as $a < b < c$.

**Fluctuational electrodynamics description of thermal emission.**

For a system of thermal objects in a nonabsorbing background reference medium, the thermally generated electric field at location $\mathbf{r}$ is defined as

$$\mathbf{E}(\mathbf{r}, \omega) = i\omega\mu_0 \int_V \bar{\bar{\mathbf{G}}}(\mathbf{r}, \mathbf{r}', \omega)\mathbf{J}^{(\text{eq})}(\mathbf{r}', \omega)\, d^3\mathbf{r}', \quad \mathbf{r} \in \Re^3, \tag{1}$$

where $i = \sqrt{-1}$, $\omega$ is the angular frequency, $\mu_0$ is the vacuum permeability, $\bar{\bar{\mathbf{G}}}(\mathbf{r}, \mathbf{r}', \omega)$ is the system Green's function, $\mathbf{J}^{(\text{eq})}(\mathbf{r}', \omega)$ is the equivalent electric current density, and integration is taken over the three-dimensional space $V$ occupied by all thermal objects and the background reference medium. The equivalent electric current density is defined as

$$\mathbf{J}^{(\text{eq})}(\mathbf{r}', \omega) = \begin{cases} \mathbf{0}, & \mathbf{r}' \in V_{\text{ref}} \\ \mathbf{J}^{(\text{fl})}(\mathbf{r}', \omega), & \mathbf{r}' \in V_{\text{therm}} \end{cases}, \tag{2}$$

where $V_{\text{therm}}$ is the volume occupied by the thermal objects, and $V_{\text{ref}}$ is the domain occupied by the nonabsorbing background reference medium characterized by a purely real dielectric function $\varepsilon_{\text{ref}}(\omega)$. The fluctuating source current density $\mathbf{J}^{(\text{fl})}(\mathbf{r}', \omega)$ arises from the thermal excitation of microcharges and is defined in terms of its autocorrelation function via the fluctuation-dissipation theorem as [62]

$$\langle \mathbf{J}^{(\text{fl})}(\mathbf{r}, \omega)\mathbf{J}^{(\text{fl})\dagger}(\mathbf{r}', \omega') \rangle = 4\pi\omega\varepsilon_0 \text{Im}[\varepsilon(\mathbf{r}, \omega)]\Theta(\omega, T)\delta(\mathbf{r} - \mathbf{r}')\delta(\omega - \omega')\bar{\bar{\mathbf{I}}}, \tag{3}$$

where $\dagger$ specifies the conjugate transpose, $\varepsilon_0$ is the vacuum permittivity, and $\delta$ is the Dirac delta function. The mean energy of an electromagnetic state at temperature $T$ and frequency $\omega$ is given by $\Theta(\omega,T)=\hbar\omega\left(e^{\frac{\hbar\omega}{k_BT}}-1\right)^{-1}$, where $\hbar$ is the reduced Planck constant and $k_B$ is the Boltzmann constant.

The system Green's function $\overline{\overline{\mathbf{G}}}(\mathbf{r},\mathbf{r}',\omega)$ in Eq. (1) is defined from the self-consistent Green's function equation [58]

$$\overline{\overline{\mathbf{G}}}^0(\mathbf{r},\mathbf{r}',\omega)=\overline{\overline{\mathbf{G}}}(\mathbf{r},\mathbf{r}',\omega)-k_0^2\int_{V_{\text{therm}}}\overline{\overline{\mathbf{G}}}^0(\mathbf{r},\mathbf{r}'',\omega)\varepsilon_{\text{r}}(\mathbf{r}'',\omega)\overline{\overline{\mathbf{G}}}(\mathbf{r}'',\mathbf{r}',\omega)\,d^3\mathbf{r}'', \tag{4}$$

where $k_0=\omega\sqrt{\mu_0\varepsilon_0}$ is the magnitude of the vacuum wavevector and the relative dielectric function is expanded as $\varepsilon_{\text{r}}(\mathbf{r},\omega)=\varepsilon(\mathbf{r},\omega)-\varepsilon_{\text{ref}}(\omega)$, with $\varepsilon(\mathbf{r},\omega)$ the dielectric function of the thermal objects. The free-space Green's function $\overline{\overline{\mathbf{G}}}^0(\mathbf{r},\mathbf{r}',\omega)$ for $\mathbf{r}'\neq\mathbf{r}$ has known analytical solution

$$\overline{\overline{\mathbf{G}}}^0(\mathbf{r},\mathbf{r}',\omega)=\frac{\exp(irk_{\text{ref}})}{4\pi r}\left[\left(1-\frac{1}{(rk_{\text{ref}})^2}+\frac{i}{rk_{\text{ref}}}\right)\overline{\overline{\mathbf{I}}}-\left(1-\frac{3}{(rk_{\text{ref}})^2}+\frac{3i}{rk_{\text{ref}}}\right)(\hat{\mathbf{r}}\hat{\mathbf{r}}^\dagger)\right], \tag{5}$$

where $r=|\mathbf{r}-\mathbf{r}'|$, $\hat{\mathbf{r}}=\frac{(\mathbf{r}-\mathbf{r}')}{|\mathbf{r}-\mathbf{r}'|}$, and $k_{\text{ref}}=\omega\sqrt{\varepsilon_{\text{ref}}(\omega)\varepsilon_0\mu_0}$. When $\mathbf{r}'=\mathbf{r}$, the analytical expression for the free-space dyadic Green's function has a singularity and must be solved using principal value techniques [63,64]. In these principal value techniques, the free-space Green's function is defined with respect to a vanishingly small exclusion volume around the singularity point. An additional term must then be added to the free-space Green's function to account for depolarization by the excluded volume. As we shall see, it is this treatment of the free-space Green's function at the source point $\mathbf{r}'=\mathbf{r}$ that is important in defining solutions for nonspherical dipole geometries.

**Near-field energy density calculations.**

Near-field thermal emission from metasurfaces is characterized by the spectral energy density at an observation point above the metasurface. The spectral energy density above a metasurface at location $\mathbf{r}$ within the background reference medium is defined as [65,66]

$$u(\mathbf{r},\omega) = \frac{\varepsilon_{\text{ref}}\varepsilon_0}{2}\langle \mathbf{E}(\mathbf{r},\omega)\cdot\mathbf{E}^*(\mathbf{r},\omega)\rangle + \frac{\mu_{\text{ref}}\mu_0}{2}\langle \mathbf{H}(\mathbf{r},\omega)\cdot\mathbf{H}^*(\mathbf{r},\omega)\rangle, \tag{6}$$

where · represents the dot product, the superscript * is the complex conjugate, and $\mu_{\text{ref}}$ is the relative permeability of the background reference medium. In this paper, we focus on modeling SiC, a material for which the magnetic contribution to the energy density is negligible in the dipole limit [67]. As such, only the electric contribution to the energy density given by the first term on the right-hand side of Eq. (6) is maintained in the following equations.

Substituting in the expression for the electric field given by Eq. (1) and simplifying, the energy density generated by a system of thermal objects may be expressed as

$$u(\mathbf{r},\omega) = \frac{\varepsilon_{\text{ref}}k_0^4}{\omega}\int_{V_{\text{therm}}} \text{Tr}\left\{\overline{\overline{\mathbf{G}}}(\mathbf{r},\mathbf{r}',\omega)\left[\overline{\overline{\mathbf{G}}}(\mathbf{r},\mathbf{r}',\omega)\right]^{\dagger}\right\}\text{Im}[\varepsilon(\mathbf{r}',\omega)]\Theta(\omega,T)\,d^3\mathbf{r}'. \tag{7}$$

For a single dipole or a metasurface constructed of $N$ thermal objects modeled as dipoles, Eq. (7) takes the discrete form

$$u(\mathbf{r},\omega) = \frac{\varepsilon_{\text{ref}}k_0^4}{\omega}\sum_j^N \Delta V_j \text{Tr}\left\{\overline{\overline{\mathbf{G}}}(\mathbf{r},\mathbf{r}_j,\omega)\left[\overline{\overline{\mathbf{G}}}(\mathbf{r},\mathbf{r}_j,\omega)\right]^{\dagger}\right\}\text{Im}[\varepsilon(\mathbf{r}_j,\omega)]\Theta[\omega,T(\mathbf{r}_j)], \tag{8}$$

where $\Delta V_j$ is the volume of the $j$th dipole. Here, $\overline{\overline{\mathbf{G}}}(\mathbf{r},\mathbf{r}_j,\omega)$ is the system Green's function relating the location $\mathbf{r}$ at which energy density is calculated with the center-of-mass location $\mathbf{r}_j$ of the $j$th dipole. This system Green's function $\overline{\overline{\mathbf{G}}}(\mathbf{r},\mathbf{r}_j,\omega)$ is the main defining parameter of near-field thermal emission from a system of dipoles and is found by discretizing Eq. (4) over the second location coordinate as

$$\overline{\overline{\mathbf{G}}}(\mathbf{r},\mathbf{r}_j,\omega) = \overline{\overline{\mathbf{G}}}^0(\mathbf{r},\mathbf{r}_j,\omega) + k_0^2\sum_k^N \Delta V_k \varepsilon_{\text{r}}(\mathbf{r}_k,\omega)\,\overline{\overline{\mathbf{G}}}^0(\mathbf{r},\mathbf{r}_k,\omega)\overline{\overline{\mathbf{G}}}(\mathbf{r}_k,\mathbf{r}_j,\omega). \tag{9}$$

The discretized free-space Green's function $\overline{\overline{\mathbf{G}}}^0(\mathbf{r},\mathbf{r}_j,\omega)$ found in both terms on the right-hand side of Eq. (9) may be represented analytically as [68]

$$\overline{\overline{\mathbf{G}}}^0(\mathbf{r},\mathbf{r}_j,\omega) = \frac{\exp(ir_{\mathbf{r}j}k_{\text{ref}})}{4\pi r_{\mathbf{r}j}}\left[\left(1-\frac{1}{(r_{\mathbf{r}j}k_{\text{ref}})^2}+\frac{i}{r_{\mathbf{r}j}k_{\text{ref}}}\right)\overline{\overline{\mathbf{I}}} - \left(1-\frac{3}{(r_{\mathbf{r}j}k_{\text{ref}})^2}+\frac{3i}{r_{\mathbf{r}j}k_{\text{ref}}}\right)\left(\hat{\mathbf{r}}_{\mathbf{r}j}\hat{\mathbf{r}}_{\mathbf{r}j}^{\dagger}\right)\right], \tag{10}$$

where $r_{\mathbf{r}j} = |\mathbf{r} - \mathbf{r}_j|$ and $\hat{\mathbf{r}}_{\mathbf{r}j} = \frac{(\mathbf{r}-\mathbf{r}_j)}{|\mathbf{r}-\mathbf{r}_j|}$. Eq. (10) does not have any singularities since the location $\mathbf{r}$ at which energy density is calculated is outside of the thermal object domain and in the background reference medium (i.e., $\mathbf{r}$ and $\mathbf{r}_j$ are never equal).

The discrete system Green's function $\bar{\bar{\mathbf{G}}}(\mathbf{r}_k, \mathbf{r}_j, \omega)$ given in the second term on the right-hand side of Eq. (9) describes the interaction among all dipoles. $\bar{\bar{\mathbf{G}}}(\mathbf{r}_k, \mathbf{r}_j, \omega)$ is calculated here using the method outlined in Ref. [58] with a special form for the self-term of the free-space Green's function to account for the ellipsoidal geometry of dipoles. To reduce computational costs in calculating $\bar{\bar{\mathbf{G}}}(\mathbf{r}_k, \mathbf{r}_j, \omega)$, we implement the weak form [69] of the free-space Green's function self-term for ellipsoidal dipoles of variable rotation, given as

$$\bar{\bar{\mathbf{G}}}^0(\mathbf{r}_i, \mathbf{r}_i, \omega) = -\frac{\mathbf{A}^{-1}\bar{\bar{\mathbf{L}}}\mathbf{A}}{\Delta V_i k_{\mathrm{ref}}^2}, \tag{11}$$

where $\mathbf{r}_i$ is the location of the center of mass of the $i$th dipole. The total rotation matrix $\mathbf{A} = \mathbf{R}_x\mathbf{R}_y\mathbf{R}_z$ accounts for rotation of the ellipsoidal dipoles from their local coordinate system by angles $\theta_x$, $\theta_y$, and $\theta_z$ around the $x$-, $y$-, and $z$-axes, respectively (see Fig. 1(a)). The individual rotation matrices $\mathbf{R}_x$, $\mathbf{R}_y$, and $\mathbf{R}_z$ are expanded as

$$\mathbf{R}_x = \begin{bmatrix} 1 & 0 & 0 \\ 0 & \cos\theta_x & -\sin\theta_x \\ 0 & \sin\theta_x & \cos\theta_x \end{bmatrix}, \tag{12}$$

$$\mathbf{R}_y = \begin{bmatrix} \cos\theta_y & 0 & \sin\theta_y \\ 0 & 1 & 0 \\ -\sin\theta_y & 0 & \cos\theta_y \end{bmatrix}, \tag{13}$$

$$\mathbf{R}_z = \begin{bmatrix} \cos\theta_z & -\sin\theta_z & 0 \\ \sin\theta_z & \cos\theta_z & 0 \\ 0 & 0 & 1 \end{bmatrix}. \tag{14}$$

The dyad $\bar{\bar{\mathbf{L}}} = \begin{bmatrix} L_1 & 0 & 0 \\ 0 & L_2 & 0 \\ 0 & 0 & L_3 \end{bmatrix}$ accounts for the depolarization of ellipsoidal dipoles and has components [64,70]:

$$L_1 = \frac{abc}{2}\int_0^\infty (a^2+q)^{-1}[(q+a^2)(q+b^2)(q+c^2)]^{-1/2}dq, \quad (15)$$

$$L_2 = \frac{abc}{2}\int_0^\infty (b^2+q)^{-1}[(q+a^2)(q+b^2)(q+c^2)]^{-1/2}dq, \quad (16)$$

$$L_3 = \frac{abc}{2}\int_0^\infty (c^2+q)^{-1}[(q+a^2)(q+b^2)(q+c^2)]^{-1/2}dq. \quad (17)$$

These geometrical factors $L_1$, $L_2$, and $L_3$ are used in calculating the polarizability tensor of ellipsoidal dipoles in standard light-scattering theory [70]. As such, Eqs. (15)–(17) may be applied to define the polarizability of ellipsoidal dipoles in many-body models of near-field radiative heat transfer.

## RESULTS AND DISCUSSION

### System description.

We model the near-field spectral energy density above single ellipsoidal particles and above metasurfaces formed from a 25-by-25 array of ellipsoidal particles. Energy density is calculated at a distance $d_{\text{obs}}$ along the vertical $z$-axis from the center of mass of single-particle systems and above the center of mass of the central particle in metasurfaces (see Fig. 1). For the metasurfaces, we model two different center-of-mass interparticle spacing values that are consistent with the assumptions for dipole approximations: $d = 3L_{\text{ch}}$ (main manuscript) and $d = 6L_{\text{ch}}$ (Supporting Information, Secs. S4, S6, S8).

All particles are made of SiC, supporting LSPhs in the infrared spectral band. The particles are embedded in a nonabsorbing medium with dielectric function $\varepsilon_{\text{ref}} = 3$. The value of $\varepsilon_{\text{ref}} = 3$ was used because it leads to increased near-field thermal energy density from SiC particles as compared with SiC particles embedded in vacuum [50]. The particle interactions with any interfaces of the host medium are neglected. The dielectric function of SiC is calculated using a Lorentz oscillator model (see Supporting Information, Sec. S1).

Each particle in a system is set to room temperature ($T = 300$ K) and is of volume $\Delta V_j = (4/3)\pi R_{\text{eq}}^3$, where $R_{\text{eq}} = 35$ nm is the radius of a sphere of equivalent volume. All particles are restricted to radial values 10 nm $\leq \{a, b, c\} \leq$ 100 nm. This range was chosen because the weak form of the free-space Green's function given in Eq. (11) does not deviate significantly from the more accurate strong form for these particle sizes. Therefore, we may model thermal emission using the weak form of the free-space Green's function, thereby decreasing computational loads without loss of accuracy.

**Single particles of variable dimensions.**

To act as a reference for subsequent metasurface calculations, we first model the energy density above single ellipsoidal particles of variable dimensions. Sphericity $\Psi$ is used to quantify the dimensions of an ellipsoid by a single representative value. Sphericity is defined as the ratio of the surface area of the ellipsoidal particle $A_{\text{ellipsoid}}$ to the surface area of a sphere of equivalent volume $A_{\text{sphere}}$,

$$\Psi = \frac{A_{\text{ellipsoid}}}{A_{\text{sphere}}}. \tag{18}$$

For a given sphericity value, there are $3! = 6$ possible ellipsoid realizations that arise from permutation of the three distinct ellipsoid semiaxes $a$, $b$, and $c$.

For these single-particle simulations, we calculate the spectral energy density at a distance $d_{\text{obs}} = 2L_{\text{ch}}$ above the center of mass of each ellipsoidal particle. Five different sphericity values are considered for ellipsoidal particles, $\Psi = 0.54083, 0.69050, 0.83094, 0.91693, 0.98926$ (Fig. 2(a)–(e)), and the resulting spectral energy densities are compared with that of a perfectly spherical particle of $\Psi = 1$ (Fig. 2(f)). For reference, we also calculate the total energy density above 55 different particles of unique sphericity ranging from $0.54083 \leq \Psi \leq 1$ (see Supporting Information, Sec. S2). The lower sphericity limit $\Psi = 0.54083$ arises from the condition that all particles are of the same volume and the requirement that the semiaxis lengths are in the range 10 nm $\leq \{a, b, c\} \leq$ 100 nm.

For the six geometric permutations of particles at a given sphericity value, there are only three unique spectra of energy density at the chosen observation point (Fig. 2(a)–(e)). The degeneracy of the spectra arises from geometric permutations that correspond to a rotation around the $z$-axis (e.g., $\{a, b, c\}$ and $\{b, a, c\}$). This degeneracy is expected to disappear when energy density is calculated at a point that does not lie along an axis of symmetry for the ellipsoidal particle.

The three resonances in the spectral energy density of ellipsoidal particles arise from shape-dependent LSPhs [70]. LSPhs exist in the Reststrahlen spectral band in which the real part of the dielectric function is negative. For SiC, the Reststrahlen band is defined between the transverse optical and longitudinal optical phonon frequencies, $\omega_{\mathrm{TO}} = 1.494 \times 10^{14}$ rad/s and $\omega_{\mathrm{LO}} = 1.824 \times 10^{14}$ rad/s, respectively. Tuning the thermal energy density of SiC particles through geometric alteration is restricted to this frequency range. For ellipsoidal particles, the three unique LSPh resonances correspond to the three unique semiaxes $a$, $b$, and $c$ that define the geometry of an ellipsoid surface. In general, the number of LSPh resonances decrease with increasing particle symmetry. For instance, spheroids have two unique semiaxis and therefore support two unique LSPh resonances, whereas spheres only have one unique radial value and support one unique LSPh resonance. The exact location, magnitude, and width of these resonances, however, are more complicated functions of material parameters.

In addition to particle shape, the frequencies at which LSPh resonances occur ($\omega_{\mathrm{LSPh}}$) are dependent on the dielectric function of the particles and the dielectric function of the background reference medium and can be approximated from the relation [70]

$$\varepsilon(\omega_{\mathrm{LSPh}}) = \varepsilon_{\mathrm{ref}}\left(1 - \frac{1}{L_i}\right), \tag{19}$$

where $L_i$ refers to the geometrical factors given in Eqs. (15)–(17). This relation is valid for particles in the dipole limit and is derived from standard electromagnetic scattering theory by determining the poles in the polarizability tensor function of ellipsoidal dipoles. Prediction of resonance frequencies for ellipsoidal particles of variable sphericity is given in Sec. S3 of the Supporting Information through solution of Eq. (19). These predicted frequencies agree with the location of resonances in the spectral energy density of the corresponding single-particle systems presented in Fig. 2(a)–(f), illustrating that near-field thermal emission is dominated by LSPhs.

In these single-particle systems, the range over which the LSPh resonances occur tends to shrink as particle sphericity increases, eventually becoming a single resonance for spherical particles (see Supporting Information, Sec. S3, Fig. S3). The same trends are seen in the spectral energy density above single particles (Fig. 2(a)–(f)). For the least spherical particle (i.e., $\Psi =$ 0.54083, see Fig. 2(a)), the peak-to-peak range of the resonances in the spectral energy density is $2.4024 \times 10^{13}$ rad/s. This range is over 5.7 times that of the peak-to-peak range ($4.154 \times 10^{12}$ rad/s) of the ellipsoidal particle with the highest sphericity (i.e., $\Psi = 0.98926$, see Fig. 2(e)). As such, the bandwidth of near-field thermal emission may be decreased by changing the particle geometry to more spherical forms. For single particles, the application space for such tuning functionality is limited. To provide a more practical example, we next model metasurfaces composed of many ellipsoidal particles.

**Metasurfaces with constituent particles of variable dimensions.**

Next, we model six different metasurfaces with constituent particles of variable sphericity and compare the resulting spectra to that of single-particle systems. All particles within a given metasurface are of the same dimensions and orientation, and particle sphericity is chosen to be consistent with that modeled in the single-particle systems of the previous section (i.e., $\Psi =$ 0.54083, 0.69050, 0.83094, 0.91693, 0.98926, 1). While we present results for observation points at $d_{\text{obs}} = 2L_{\text{ch}}$, similar spectral behavior is seen for farther observation points tested at $d_{\text{obs}} = 7L_{\text{ch}}$ (not shown).

The spectral behavior of thermal emission from metasurfaces with close interparticle spacing (i.e., $d = 3L_{\text{ch}}$, see Fig. 3(a)–(f)) is found to deviate from that of individual constituent particles. Specifically, the maximum absolute value of the relative difference between the normalized energy density spectra of a metasurface and the corresponding individual particle (i.e., $\max\left[|\bar{u}_{\text{metasurface}}(\mathbf{r},\omega) - \bar{u}_{\text{particle}}(\mathbf{r},\omega)|/\bar{u}_{\text{particle}}(\mathbf{r},\omega)\right]$) is between 42% and 941% over the Reststrahlen band for all metasurfaces modeled. Conversely, the spectrum of thermal emission from metasurfaces with larger interparticle spacing (i.e., $d = 6L_{\text{ch}}$, see Supporting Information, Sec. S4, Fig. S5(a)–(f)) is found to be approximated to a high degree of accuracy by that of individual constituent particles (i.e., $\max\left[|\bar{u}_{\text{metasurface}}(\mathbf{r},\omega) - \bar{u}_{\text{particle}}(\mathbf{r},\omega)|/\bar{u}_{\text{particle}}(\mathbf{r},\omega)\right] \lesssim 21\%$ over the Reststrahlen band for all metasurfaces modeled). These trends have been observed for arrays of spherical dipoles [50] and may be attributed to the degree of multiparticle interaction within a metasurface [71,72]. In the metasurfaces with close interparticle spacing $d = 3L_{\text{ch}}$, multiparticle interaction significantly influences the near-field spectral energy density. This contribution may be seen from the relative value of off-diagonal terms in the discrete system Green's function $\bar{\bar{\mathbf{G}}}(\mathbf{r}_k, \mathbf{r}_j, \omega)$ that describes interaction between all particles in a metasurface (see Supporting Information, Sec. S5). For metasurfaces with close interparticle spacing $d = 3L_{\text{ch}}$, the off-diagonal terms of $\overline{\overline{\mathbf{G}}}(\mathbf{r}_k, \mathbf{r}_j, \omega)$ are relatively large, indicating significant multiparticle interaction (see Supporting Information, Sec. S5, Figs. S6(a)–(b)). As the interparticle spacing is increased, however, the discrete system Green's function $\overline{\overline{\mathbf{G}}}(\mathbf{r}_k, \mathbf{r}_j, \omega)$ becomes increasingly diagonal (see Supporting Information, Sec. S5, Figs. S6(c)–(d)), signifying that individual particle self-interaction dominates, multiparticle interactions are negligible, and the spectral energy density of the metasurface may be approximated as the linear combination of spectral energy density values of constituent particles.

Similar analysis may be used to explain why the spectra of metasurfaces composed of less spherical particles (i.e., small $\Psi$) more closely match the corresponding single-particle spectra even when the metasurface has closer interparticle spacing $d = 3L_{\text{ch}}$ (Fig. 3(a)–(f)). This trend can be seen explicitly when comparing the spectra of the highest sphericity ($\Psi = 1$, Fig. 3(f)) and lowest sphericity ($\Psi = 0.54083$, Fig. 3(a)) systems. The maximum relative difference in the spectra of metasurfaces versus single particles with $d = 3L_{\text{ch}}$ and $\Psi = 1$ (Fig. 3(f)) is 941%, whereas this value is only 83% for metasurfaces and single particles with $d = 3L_{\text{ch}}$ and $\Psi = 0.54083$ (Fig. 3(a)). Since the interparticle spacing is defined with respect to the characteristic length of constituent particles (i.e., $d = 3L_{\text{ch}}$) and less spherical particles have greater characteristic length values, the lower sphericity metasurfaces correspond to systems with larger interparticle spacing. As such, multiparticle interaction in these systems is less significant, and the spectra of these metasurfaces may be approximated by that of single particles.

**Metasurfaces with constituent particles of variable orientation.**

Next, we model metasurfaces for which the constituent ellipsoidal particles are all rotated by the same angle $\theta_y$ around the $y$-axis in the local coordinate system of every particle (Fig. 4). All local coordinate systems are parallel to the global Cartesian coordinate system. In the metasurface modeled, the constituent ellipsoidal particles are all the same dimensions (i.e., $a =$ 10.06 nm, $b =$ 56.62 nm, $c =$ 75.24 nm, $\Psi =$ 0.54083), and interparticle spacing is $d = 3L_{\text{ch}}$. Energy density is calculated at a distance $d_{\text{obs}} = 2L_{\text{ch}}$ above the center of mass of the central particle. The spectral profile at the observation point $d_{\text{obs}} = 2L_{\text{ch}}$ is consistent with the spectra at farther separation distances (e.g., $d_{\text{obs}} = 7L_{\text{ch}}$, not shown), with only differences in magnitude.

Rotation of all particles in the array results in damping of the low-frequency resonance at $1.554 \times 10^{14}$ rad/s, amplification of the high-frequency resonance at $1.794 \times 10^{14}$ rad/s, and minimal change in the middle resonance at $1.579 \times 10^{14}$ rad/s (Fig. 4(a)). There is negligible spectral shift in resonances with change in ellipsoid orientation. This independence of particle orientation and resonance location is expected in the regime in which multiparticle interaction effects are less significant. Based on the results in Fig. 3(a) that compare an equivalent unrotated metasurface with a single particle, we can assume that we are in this regime where individual particle self-interaction dominates. While different in magnitude, the spectral profile of energy density for the metasurfaces with particles at $\theta_y = 0$ rad (black line, Fig. 4(a)) and $\theta_y = \pi/2$ rad/s (blue line, Fig. 4(a)) match, respectively, the single-particle spectra for $\{a, b, c\}$ and $\{c, b, a\}$ particles of sphericity $\Psi =$ 0.54083 in Fig. 2(a). These single-particle geometric permutations correspond to the same orientation as that of the constituent particles in the metasurface at, respectively, $\theta_y = 0$ rad and $\theta_y = \pi/2$ rad/s. As such, the spectral energy density of each of these metamaterials with similarly oriented particles should be well represented by a linear combination of single-particle spectra. The same conclusions are applicable when interparticle spacing is $d = 6L_{\text{ch}}$ (see Supporting Information, Sec. S6).

The total, spectrally integrated energy density at $\theta_y = 0$ rad is almost double that at $\theta_y = \pi/2$ rad/s (see Supporting Information, Sec. S7). As such, metasurfaces like this one with dynamic control of the orientation of constituent particles could be implemented as thermal switches.

**Metasurfaces with randomized constituent particles.**

Next, we introduce randomization into metasurfaces. Randomization is incorporated in three different ways: (1) random particle dimensions, (2) random particle orientation, and (3) both random particle dimensions and random particle orientation. In all systems, particle semiaxes are constrained as 10 nm $\leq \{a, b, c\} \leq$ 100 nm with $a < b < c$, and the particle characteristic length is defined as the maximum radial dimension over all particles in the metasurface: $L_{\text{ch}} = \max(\{\{a_i\}, \{b_i\}, \{c_i\}\})$. Metasurfaces with interparticle spacing of $d = 3L_{\text{ch}}$ (Fig. 5) and $d = 6L_{\text{ch}}$ (Supporting Information, Sec. S8) are modeled. The spectral energy density is calculated at six different distances above the central particle in each metasurface, from $d_{\text{obs}} = 2L_{\text{ch}}$ to $7L_{\text{ch}}$.

In the random particle dimensions metasurfaces (Figs. 5(a)–(b) and Supporting Information, Sec. S8, Fig. S9(a)–(b)), particles are oriented as $\theta_x = \theta_y = \theta_z = 0$. The spectral energy density of these metasurface differs noticeably from all previous cases. Instead of three distinct resonances, the spectral energy density displays multiple resonances and becomes increasingly broadband at farther distances above the metasurface. Focusing on the metasurface with interparticle spacing $d = 3L_{\text{ch}}$, the spectral energy density has five main resonances at the observation point $d_{\text{obs}} = 2L_{\text{ch}}$. In this case, the location of the three dominant resonances may be predicted from the geometric parameters of the central particle in the 25-by-25 array above which the observation point is located. This spectral behavior is due to dominance of the central particle in the array at very close observation points in addition to only minor multiparticle interaction effects. At farther observation points, the central particle in the metasurface becomes less dominant to thermal emission as the same solid angle from the observation point encompasses a larger number of particles. This same logic may be applied in interpreting the spectral energy density above the metasurface with interparticle spacing $d = 6L_{\mathrm{ch}}$. This implies that it is critical to control particle dimensions to obtain narrowband thermal emission from metasurfaces made of dielectric particles. In all of these cases, the spectral energy density may be approximated as the summation of all single-particle spectra weighted by the distance from the individual particle to the observation point. This approximation will begin to deviate from actual thermal emission spectra when interparticle spacing is small enough that multiparticle interactions become important.

Next, we model a metasurface composed of randomly orientated particles of low-sphericity ellipsoids (i.e., $\Psi = 0.54083$) and random rotation angles $\theta_x$, $\theta_y$, and $\theta_z$ for each particle (Fig. 5(c)–(d) and Supporting Information, Sec. S8, Fig. S9(c)–(d)). As expected from the analysis in the previous section, randomizing the orientation of constituent particles only results in a change in the magnitude of each resonance and does not affect the frequency location of resonances in the spectral energy density. At all observation points, the spectral energy density displays three clear resonances at the same frequencies as LSPh resonances of single-particle systems and non-randomized metasurfaces constructed of particles of the same dimensions. Multiparticle interaction effects are inferred to be less significant. These results imply that tunable, narrowband thermal emission can be achieved from metasurfaces made of dielectric particles dominated by individual particle self-interaction effects if the dimensions of constituent particles are well-controlled, irrespective of particle alignment.

Finally, we model a metasurface composed of particles with random dimensions and random orientations (Fig. 5(e)–(f) and Supporting Information, Sec. S8, Fig. S9(e)–(f)). The spectral energy density displays similar trends to that of the metasurface made of unrotated, randomly dimensioned particles: the spectral energy density from the metasurface can be approximated as a weighted sum of single-particle spectra, and the spectral energy density becomes increasingly broadband over the Reststrahlen band for farther observation points. These results lead us to expect that actual metasurfaces made of dielectric particles manufactured with non-perfect particle alignment and constructed of particles defined by size distributions rather than exact dimensions will display an averaging of LSPh resonances and more broadband spectra than idealized structures.

**CONCLUSIONS**

We have presented an exact method for calculating near-field energy density from metasurfaces composed of ellipsoidal particles in the dipole limit. This method is derived using a variation of the discrete system Green's function method and can be applied to resolve unique trends in the spectral energy density at arbitrary observation points above particle-based metasurfaces. We applied this method to model a variety of metasurfaces composed of SiC ellipsoidal particles of variable dimensions and orientation. The main finding of this work is that the geometric parameters of constituent particles can be used to control the frequency and bandwidth of LSPh resonances, thereby providing tunability of the spectral energy density from dielectric particle-based metasurfaces. Metasurfaces with interparticle spacing larger than a few times the characteristic length of the constituent particles are dominated by individual particle self-interaction. As such, the spectral thermal emission of these metasurfaces may be approximated as a linear combination of individual-particle spectra. For metasurfaces with interparticle spacing below this limit, multiparticle interactions come into play, and full-system models are required to accurately resolve the spectral energy density. These results are important for characterizing dielectric particle-based metasurfaces that are constructed of particles defined by distributions of geometric parameters rather than exact values.

## ASSOCIATED CONTENT

The Supporting Information is available free of charge at [URL to be included by the publisher].

Dielectric function of SiC; spectrally integrated energy density results; frequency prediction of LSPh resonances for ellipsoidal dipoles; visualization of discrete system Green's function; spectral energy density results for metasurfaces with interparticle spacing of $d = 6L_{\mathrm{ch}}$

## AUTHOR INFORMATION

### Corresponding Authors

*E-mail: L.Walter@utah.edu, mathieu.francoeur@mcgill.ca

### ORCID

Mathieu Francoeur: 0000-0003-4989-4861

Lindsay Walter: 0000-0003-3416-6079

### Author Contributions

This work was conceived by L.P.W. and M.F. Methodology was developed by L.P.W. and J.M. under the supervision of M.F. Numerical simulations were performed by L.P.W. under the supervision of M.F. The manuscript was written by L.P.W., J.M., B.R., and M.F.

### Notes

The authors declare no competing financial interest.

## CONFLICT OF INTEREST

None.

## FUNDING SOURCES

National Science Foundation, Grants No. CBET-1952210 and CMMI-2130083.

## ACKNOWLEDGEMENTS

The authors acknowledge financial support from the National Science Foundation (Grants No. CBET-1952210 and CMMI-2130083). L.P.W. acknowledges that this material is based upon work supported by the National Science Foundation Graduate Research Fellowship under Grant No. DGE-1747505. Any opinions, findings, and conclusions or recommendations expressed in this material are those of the authors and do not necessarily reflect the views of the National Science Foundation.

FIGURES

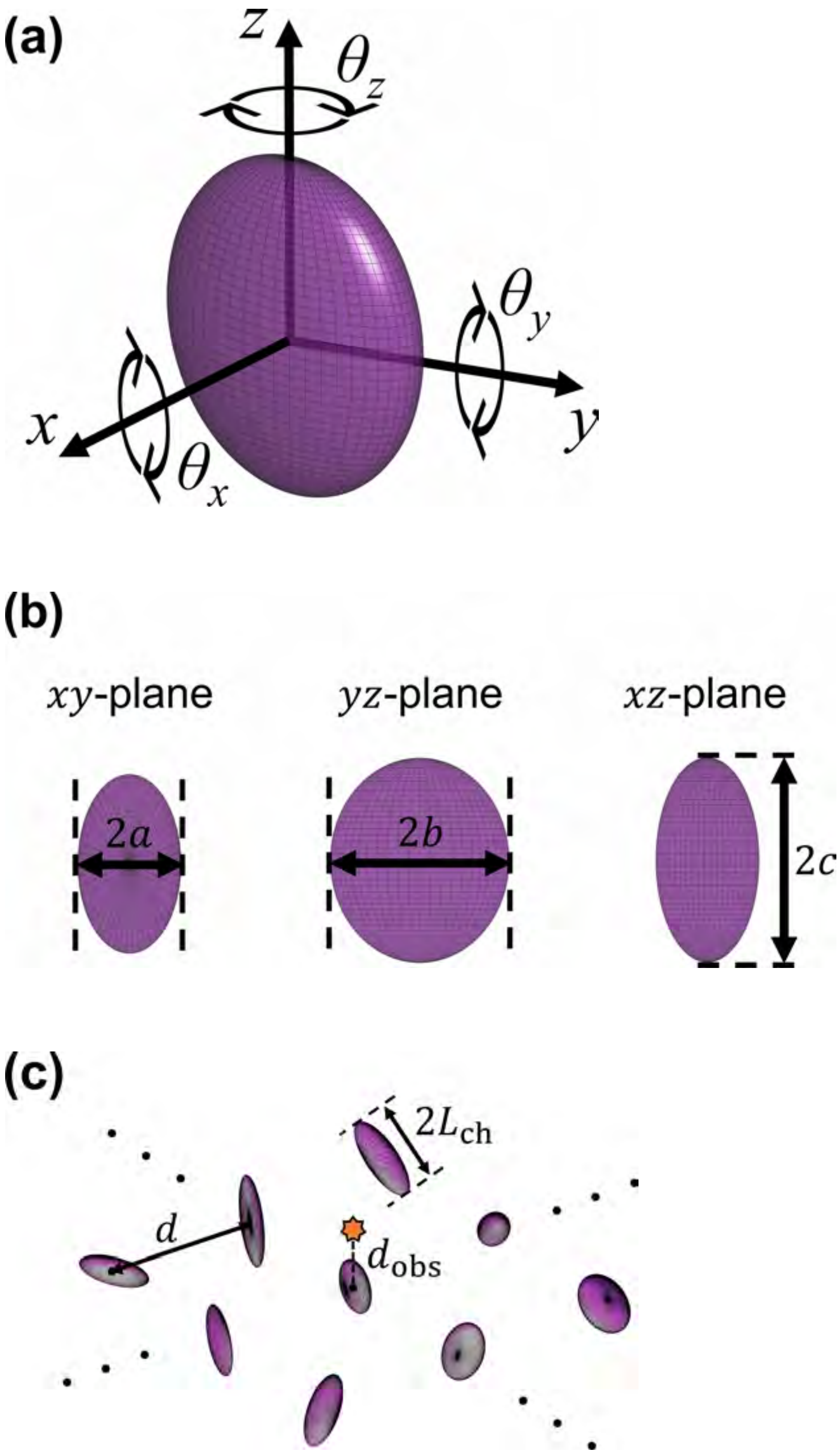

**Figure 1.** (a) Ellipsoidal particle in its local coordinate system with rotation convention depicted. (b) The three major semiaxis dimensions of the ellipsoidal particle from panel (a). (c) Metasurface composed of a single layer of ellipsoidal particles. The observation point at which energy density is calculated is depicted by the yellow star and is located directly above the central particle in the 25-by-25 array.

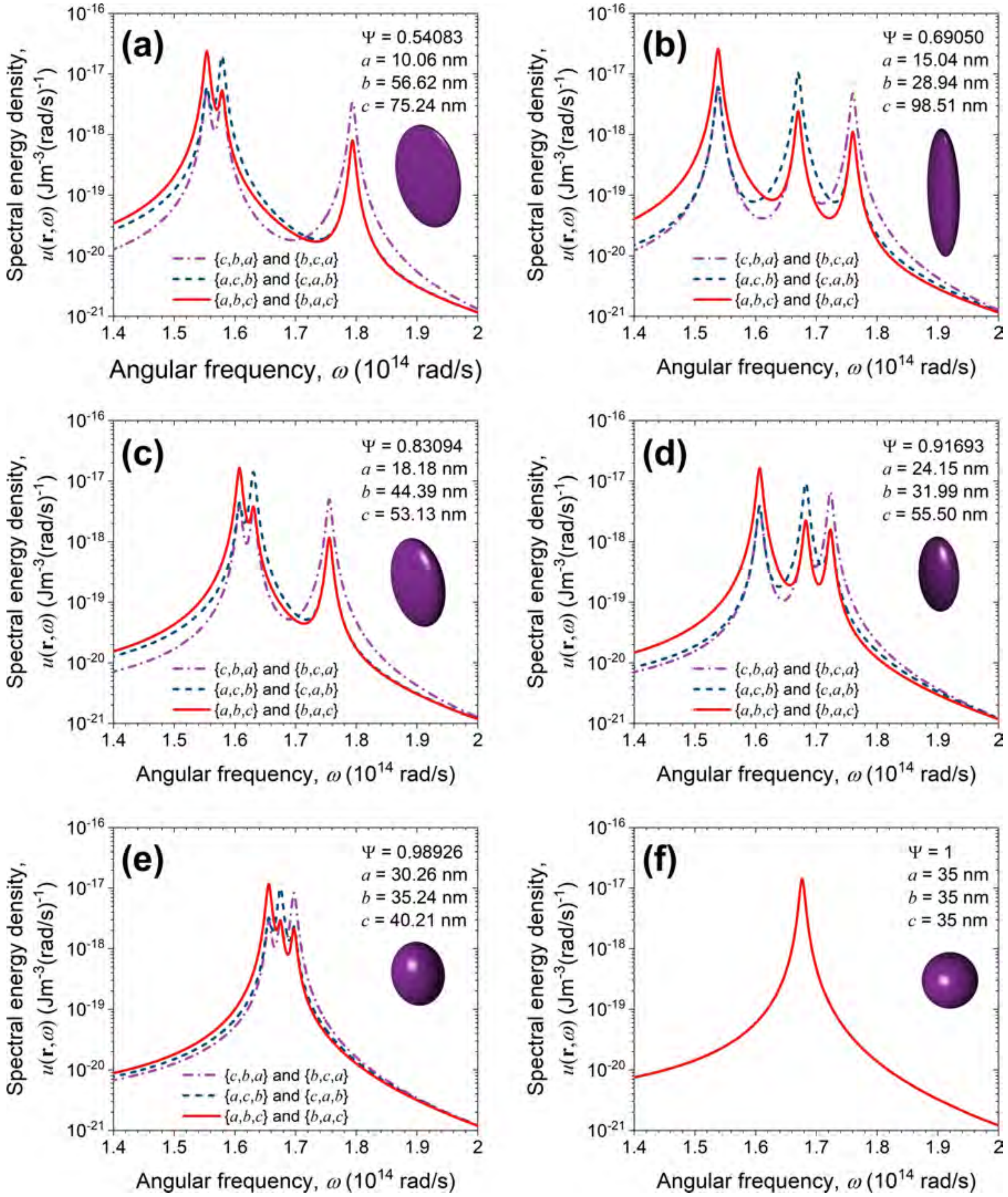

**Figure 2.** Spectral energy density above a single particle of variable sphericity. All particles are SiC, at temperature $T\ = 300$ K, and embedded in a nonabsorbing background medium of $\varepsilon_{\text{ref}} = 3$. Each sphericity value corresponds to six geometric permutations of the ellipsoidal particles, represented by sets in curly brackets (panels (a)–(e)). Particles are unrotated such that the particle center of mass is set at the origin and the three major ellipsoid semiaxes are aligned with the axes of the Cartesian coordinate system. The observation point is located at a vertical distance of $d_{\text{obs}} = 2L_{\text{ch}}$ above each particle along the $z$-axis.

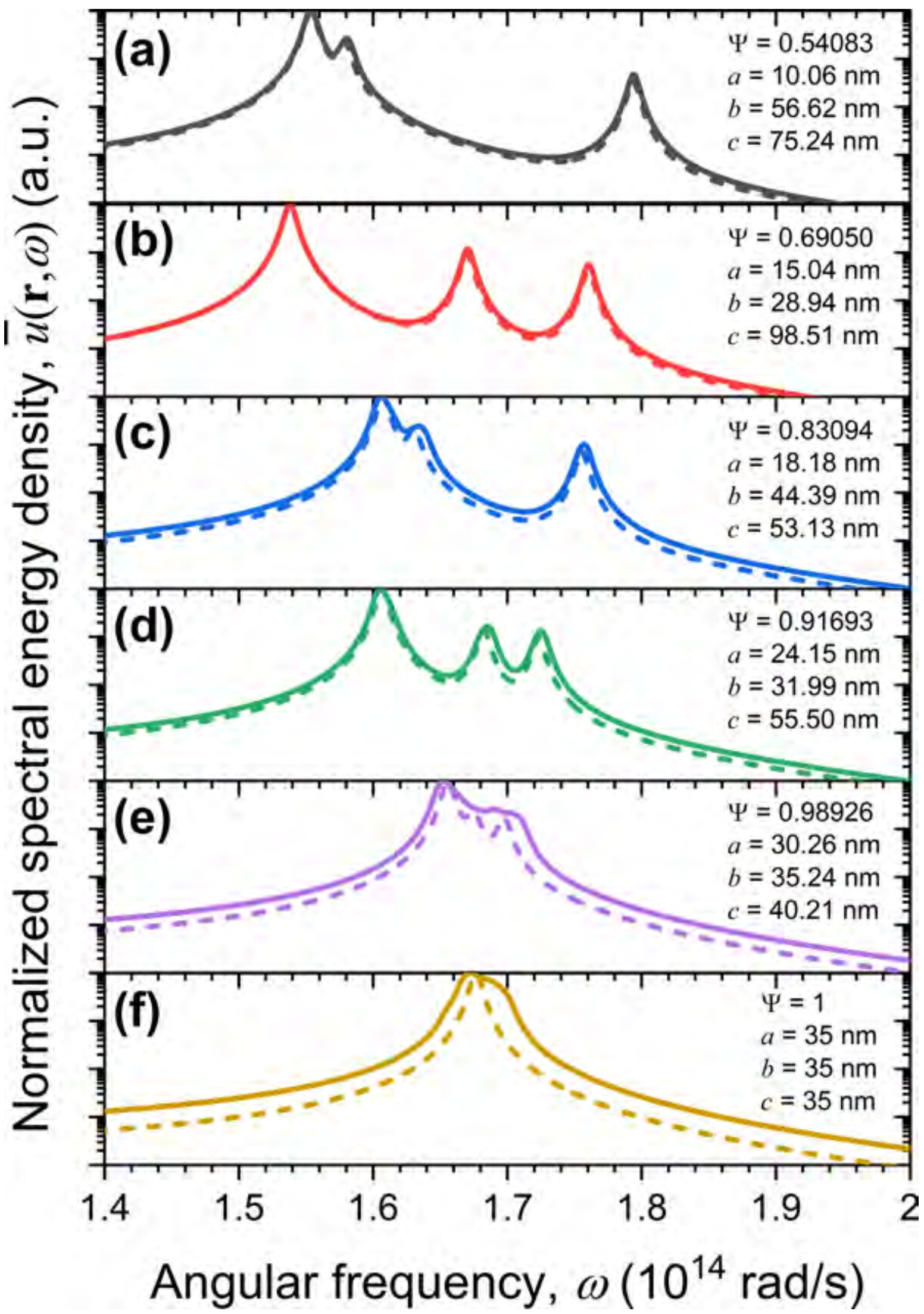

**Figure 3.** Normalized spectral energy density above metasurfaces with constituent particles of variable sphericity. Normalization is implemented as $\bar{u}(\mathbf{r},\omega) = u(\mathbf{r},\omega)/\max[u(\mathbf{r},\omega)]$ for each system. Each metasurface is composed of a 25-by-25 array of ellipsoidal SiC particles at temperature $T = 300$ K and embedded in a nonabsorbing background medium of $\varepsilon_{\mathrm{ref}} = 3$. Solid lines represent metasurfaces and dashed lines represent a single particle. The sphericity of each particle is varied from $\Psi = 0.5408$ to $\Psi = 1$. All particles are unrotated. Interparticle spacing is $d = 3L_{\mathrm{ch}}$. The observation point is located at a vertical distance of $d_{\mathrm{obs}} = 2L_{\mathrm{ch}}$ above the center of mass of the central particle in the array.

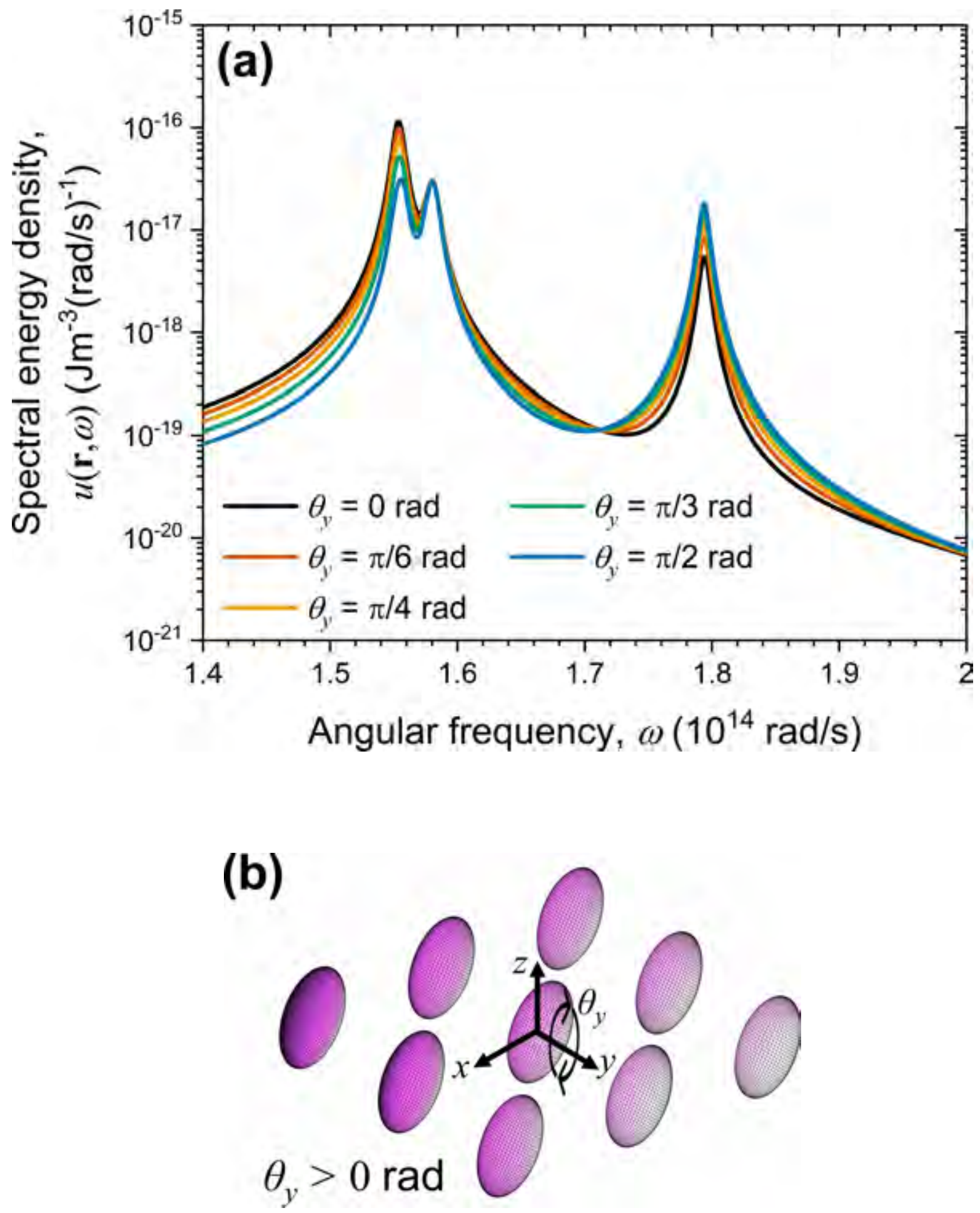

**Figure 4.** (a) Spectral energy density above metasurfaces with constituent particles of variable rotation angle. Each metasurface is composed of a 25-by-25 array of ellipsoidal SiC particles at temperature $T$ = 300 K and embedded in a nonabsorbing background medium of $\varepsilon_{\text{ref}} = 3$. (b) Particles are rotated by the same angle $\theta_y$ in their local coordinate system and are of the same dimensions ($a$ = 10.06 nm, $b$ = 56.62 nm, $c$ = 75.24 nm, $\Psi = 0.5408$). Interparticle spacing is $d = 3L_{\text{ch}}$. The observation point is located at a vertical distance of $d_{\text{obs}} = 2L_{\text{ch}}$ above the center of mass of the central particle in the array.

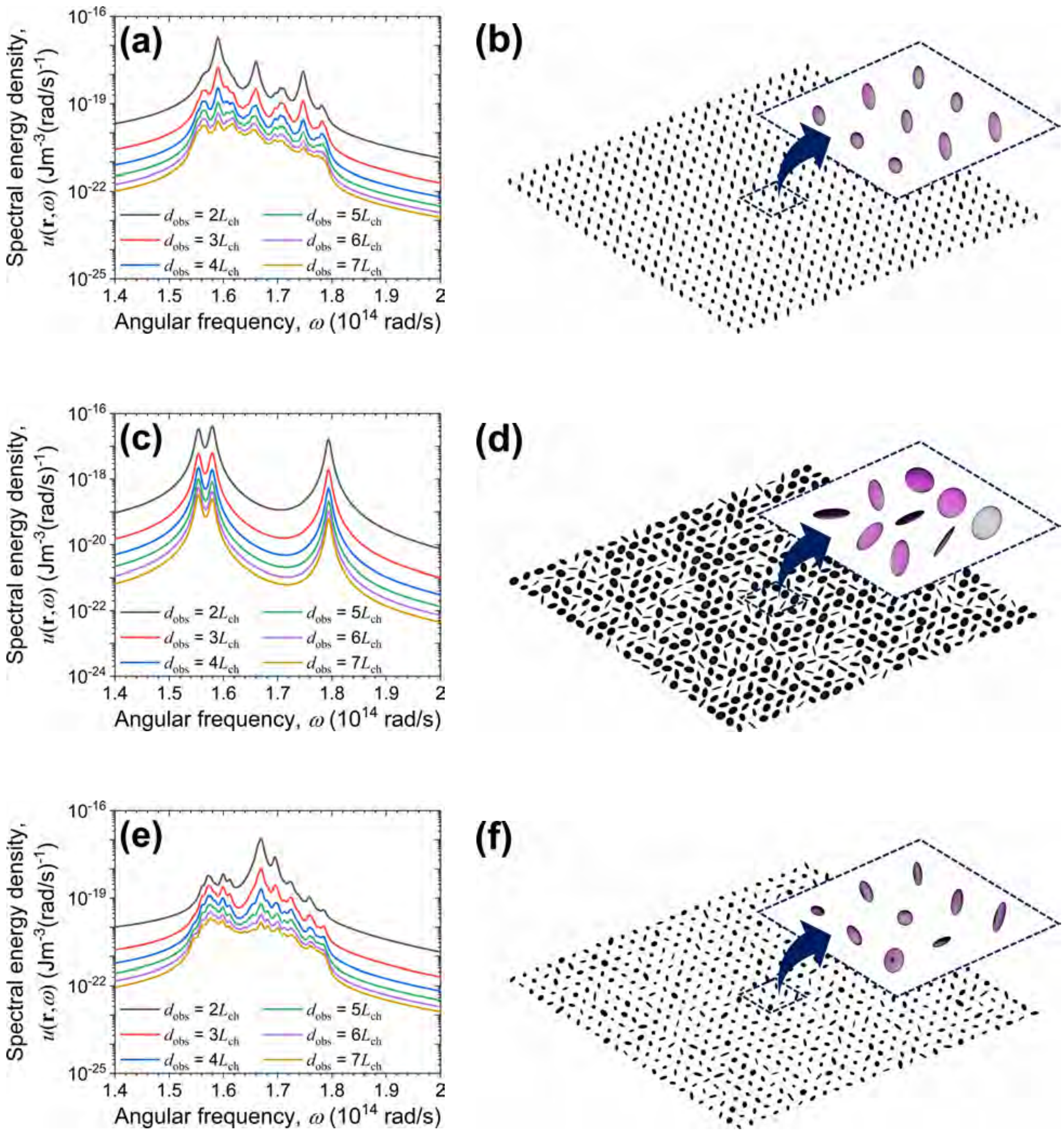

**Figure 5.** Spectral energy density above metasurfaces with randomized constituent particles. Each metasurface is composed of a 25-by-25 array of ellipsoidal SiC particles at temperature $T = 300$ K and embedded in a nonabsorbing background medium of $\varepsilon_{\text{ref}} = 3$. (a)–(b) Particles are unrotated and of random dimensions; (c)–(d) particles are all the same dimensions ($a = 10.06$ nm, $b = 56.62$ nm, $c = 75.24$ nm, $\Psi = 0.5408$) and of random orientation; and (e)–(f) particles are of random dimensions and random orientation. Interparticle spacing is $d = 3L_{\text{ch}}$. Spectral energy density is measured at six different observation points located at vertical distances of $d_{\text{obs}} = 2L_{\text{ch}}$, $3L_{\text{ch}}$, $4L_{\text{ch}}$, $5L_{\text{ch}}$, $6L_{\text{ch}}$, $7L_{\text{ch}}$ above the center of mass of the central particle in the array.

# Supporting Information

## Near-field thermal emission from metasurfaces constructed of SiC ellipsoidal particles

Lindsay P. Walter[†,*], Joseph C. McKay[†], Bart Raeymaekers[‡], and Mathieu Francoeur[‡‡,*]

[†]Radiative Energy Transfer Lab, Department of Mechanical Engineering, University of Utah, Salt Lake City, UT 84112, USA

[‡]Manufacturing and Tribology Laboratory, Department of Mechanical Engineering, Virginia Tech, Blacksburg, VA 24061, USA

[‡‡]Department of Mechanical Engineering, McGill University, Montréal, QC H3A 0C3, Canada

[*]Corresponding authors: L.Walter@utah.edu, mathieu.francoeur@mcgill.ca

## Table of Contents:

## S1. DIELECTRIC FUNCTION OF SiC

The dielectric function of SiC used in the simulations is modeled using a Lorentz oscillator model with parameters given in Ref. [1] (Fig. S1). The equation for the dielectric function and relevant parameters are presented in Eq. (S1) and Table SI, respectively,

$$\varepsilon(\omega) = \varepsilon_\infty \left(\frac{\omega^2-\omega_{\mathrm{LO}}^2+i\Gamma\omega}{\omega^2-\omega_{\mathrm{TO}}^2+i\Gamma\omega}\right). \quad \text{(S1)}$$

**TABLE SI.** Parameters for SiC dielectric function calculation.

| Description | Parameter | Value | Units |
|---|---|---|---|
| High-frequency permittivity limit | $\varepsilon_\infty$ | 6.7 | – |
| Longitudinal optical phonon frequency | $\omega_{\mathrm{LO}}$ | $1.825 \times 10^{14}$ | rad/s |
| Transverse optical phonon frequency | $\omega_{\mathrm{TO}}$ | $1.494 \times 10^{14}$ | rad/s |
| Damping constant | $\Gamma$ | $8.966 \times 10^{11}$ | rad/s |

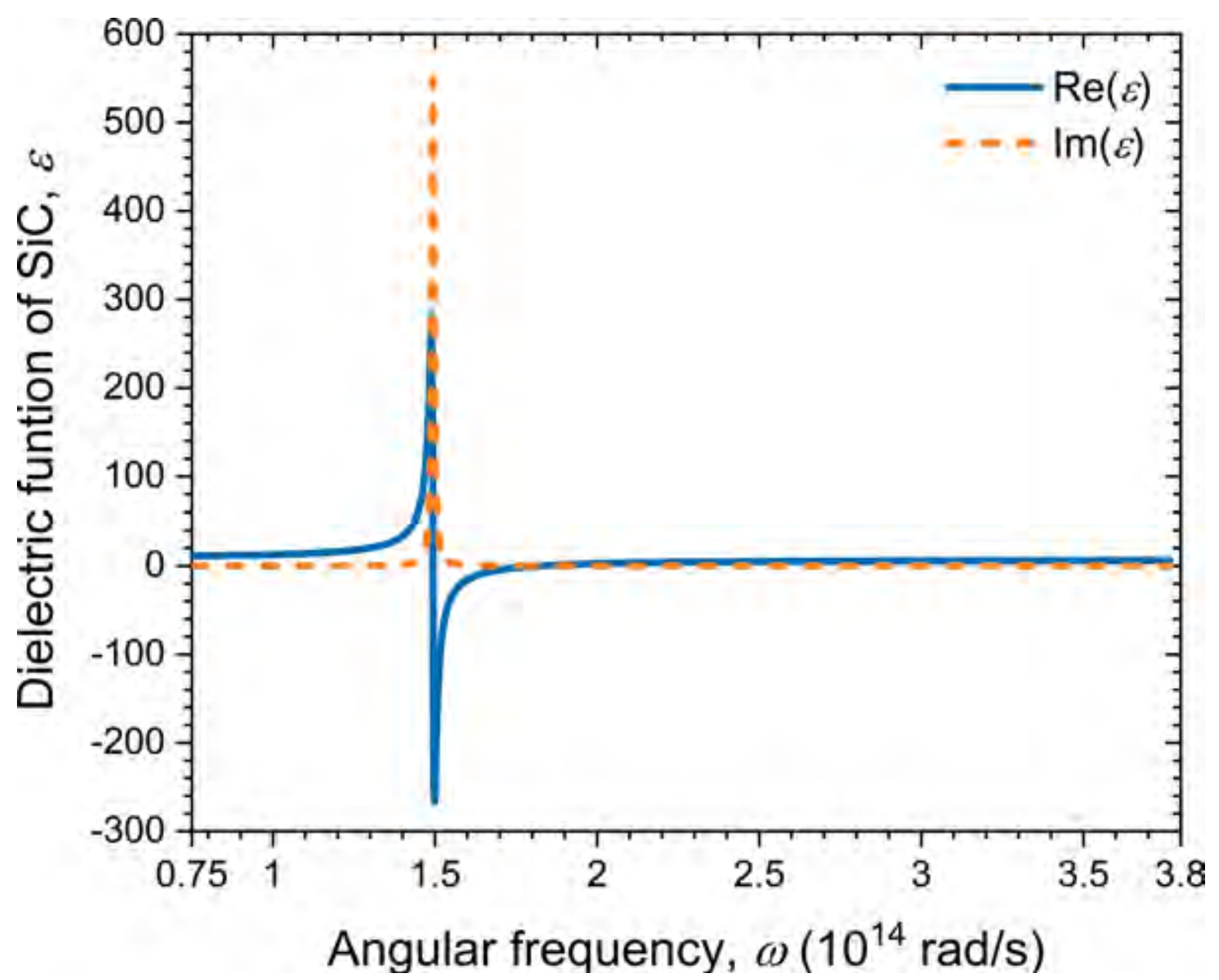

**Figure S1.** Dielectric function of SiC defined using a Lorentz oscillator model. A total of 1200 frequencies within the spectral band $7.5 \times 10^{13}$ rad/s to $3.8 \times 10^{14}$ rad/s are implemented for calculations of the total, spectrally integrated energy density.

## S2. TOTAL ENERGY DENSITY ABOVE SINGLE SiC PARTICLES OF VARIABLE SPHERICITY

The total (i.e., spectrally integrated) energy density above single particles at an observation point $d_{\text{obs}} = 2L_{\text{ch}}$ is presented in Fig. S2. Here, total energy density values are given for particles with 55 unique sphericity values ranging from $0.54083 \leq \Psi \leq 1$ and include data points for the particles presented in Fig. 2 of the main manuscript. Table SII lists the particle dimensions that correspond to data points in Fig. S2.

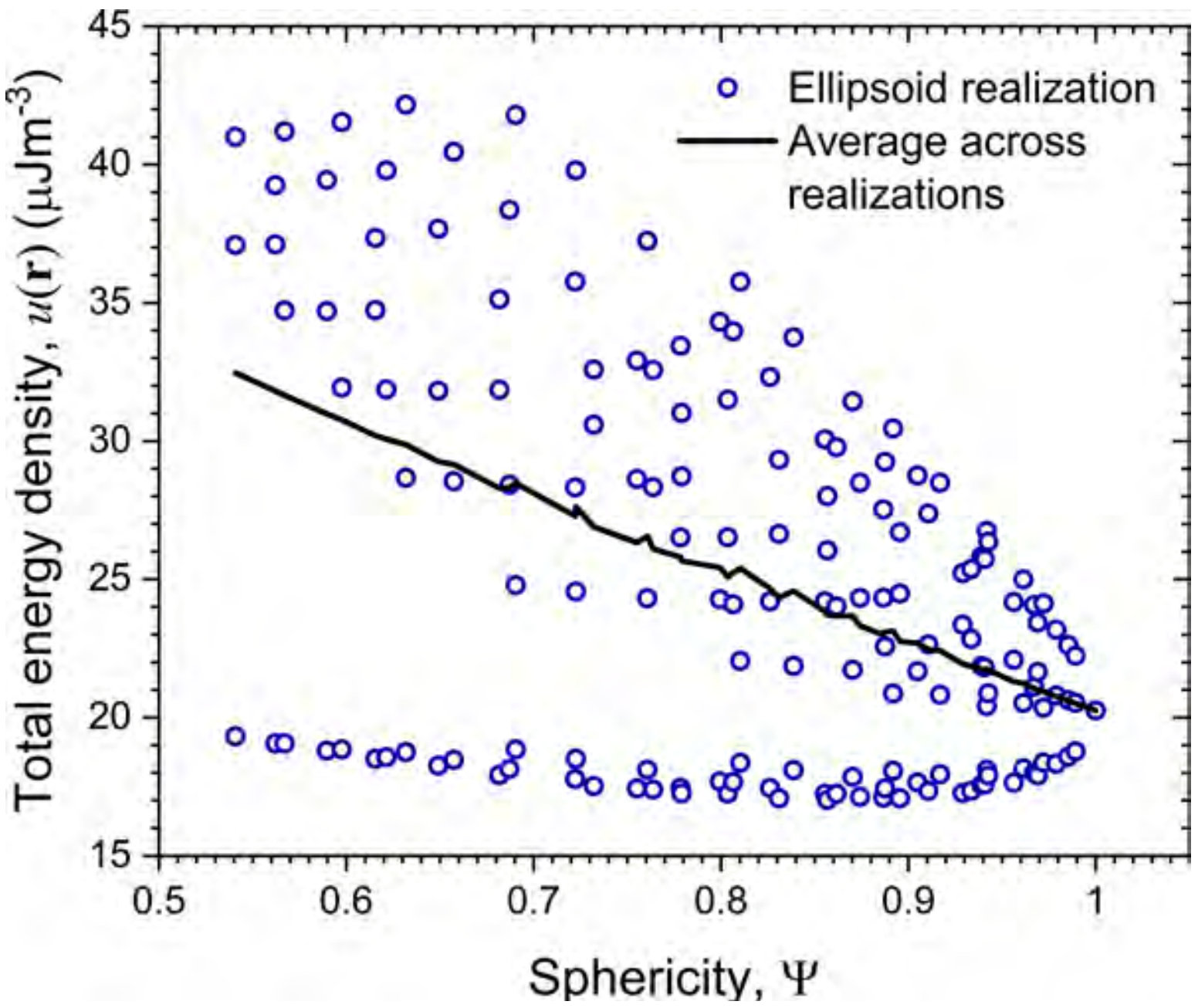

**Figure S2.** Total energy density above a single particle of variable sphericity. Each sphericity value corresponds to six geometric permutations of an ellipsoidal particle, and the solid black line represents the average of these six permutations. All particles are SiC, at temperature $T = 300$ K, and embedded in a nonabsorbing background medium of $\varepsilon_{\text{ref}} = 3$. Particles are unrotated such that the particle center of mass is set at the origin and the three major ellipsoid semiaxes are aligned with the axes of the Cartesian coordinate system. The observation point is located at a vertical distance of $d_{\text{obs}} = 2L_{\text{ch}}$ above each particle along the $z$-axis.

**TABLE SII.** Ellipsoid dimensions that correspond to each sphericity value presented in Fig. S2. The ellipsoid geometries that are modeled in the main manuscript are highlighted in gray.

| Sphericity, Ψ | Ellipsoid semiaxis dimensions | | |
|---|---|---|---|
| | $a$ (nm) | $b$ (nm) | $c$ (nm) |
| 0.54083 | 10.06379 | 56.62060 | 75.24332 |
| 0.56214 | 10.51685 | 59.16957 | 68.90012 |
| 0.56707 | 10.68492 | 50.22911 | 79.88725 |
| 0.58954 | 11.16594 | 52.49035 | 73.15255 |
| 0.59759 | 11.49477 | 43.40080 | 85.94217 |
| 0.61557 | 11.74724 | 55.22299 | 66.09193 |
| 0.62145 | 12.01224 | 45.35464 | 78.69703 |
| 0.63187 | 12.62391 | 35.98405 | 94.38439 |
| 0.64915 | 12.63760 | 47.71580 | 71.10126 |
| 0.65740 | 13.19222 | 37.60400 | 86.42755 |
| 0.68165 | 13.41758 | 50.66076 | 63.07515 |
| 0.68715 | 13.87901 | 39.56166 | 78.08564 |
| 0.69050 | 15.03588 | 28.94758 | 98.50608 |
| 0.72222 | 14.73560 | 42.00336 | 69.27111 |
| 0.72252 | 15.81865 | 30.45459 | 88.99837 |
| 0.73209 | 14.80392 | 50.37646 | 57.49097 |
| 0.75511 | 15.57462 | 45.51417 | 60.48395 |
| 0.76063 | 16.79495 | 32.33421 | 78.95198 |
| 0.76385 | 15.85247 | 45.18694 | 59.85417 |
| 0.77849 | 16.53586 | 40.37640 | 64.21694 |
| 0.77895 | 16.27576 | 47.56315 | 55.38500 |
| 0.79943 | 17.78917 | 34.88750 | 69.08416 |
| 0.80369 | 17.28028 | 42.19409 | 58.80329 |
| 0.80655 | 18.06790 | 34.78493 | 68.21899 |
| 0.81023 | 19.53663 | 28.92559 | 75.87040 |
| 0.82621 | 18.59001 | 36.45809 | 63.26020 |
| 0.83094 | 18.17989 | 44.39071 | 53.12765 |
| 0.83869 | 20.41614 | 30.22778 | 69.47433 |
| 0.85548 | 19.55781 | 38.35609 | 57.15438 |
| 0.85682 | 19.23086 | 44.19944 | 50.44159 |
| 0.86168 | 19.84274 | 38.20190 | 56.56107 |
| 0.87030 | 21.47900 | 31.80143 | 62.76873 |
| 0.87416 | 20.23201 | 39.93335 | 53.06758 |
| 0.88669 | 20.76490 | 40.72339 | 50.70263 |
| 0.88765 | 21.48071 | 35.42556 | 56.34284 |
| 0.89188 | 23.10880 | 30.60970 | 60.61326 |
| 0.89544 | 21.14283 | 41.73109 | 48.59385 |
| 0.90486 | 22.80466 | 33.76418 | 55.68322 |

**TABLE SII (continued)**

| | | | |
|---|---|---|---|
| 0.91067 | 22.44774 | 37.02037 | 51.59300 |
| 0.91693 | 24.14913 | 31.98770 | 55.50341 |
| 0.92901 | 23.19457 | 40.24602 | 45.92984 |
| 0.93354 | 23.61637 | 38.94765 | 46.61329 |
| 0.93908 | 24.40208 | 36.36151 | 48.32094 |
| 0.94076 | 24.53310 | 36.32328 | 48.11346 |
| 0.94187 | 25.90814 | 32.25692 | 51.30325 |
| 0.94260 | 25.40633 | 33.65298 | 50.14627 |
| 0.95631 | 25.50062 | 37.99845 | 44.24737 |
| 0.96142 | 27.07449 | 33.70908 | 46.97826 |
| 0.96699 | 26.97438 | 35.73000 | 44.48562 |
| 0.96908 | 26.84380 | 37.41055 | 42.69392 |
| 0.97223 | 28.24129 | 33.79972 | 44.91657 |
| 0.97893 | 28.48399 | 35.46398 | 42.44396 |
| 0.98538 | 29.51267 | 35.32133 | 41.12999 |
| 0.98926 | 30.25959 | 35.23583 | 40.21208 |
| 1.00000 | 35.00000 | 35.00000 | 35.00000 |

## S3. FREQUENCIES OF LOCALIZED SURFACE PHONON RESONANCES IN ELLIPSOIDAL DIPOLES

Localized surface phonon (LSPh) resonances are predicted for ellipsoidal dipoles of variable sphericity in Fig. S3. The general relation between LSPh resonance frequency and the geometrical parameter $L_i$ for ellipsoidal dipoles as given in Eq. (19) of the main manuscript is presented in Fig. S4. Here, ellipsoidal dipoles are made of SiC and embedded in a background reference medium with dielectric function $\varepsilon_{\mathrm{ref}} = 3$. The LSPh resonance frequencies will be different for systems with different optical properties.

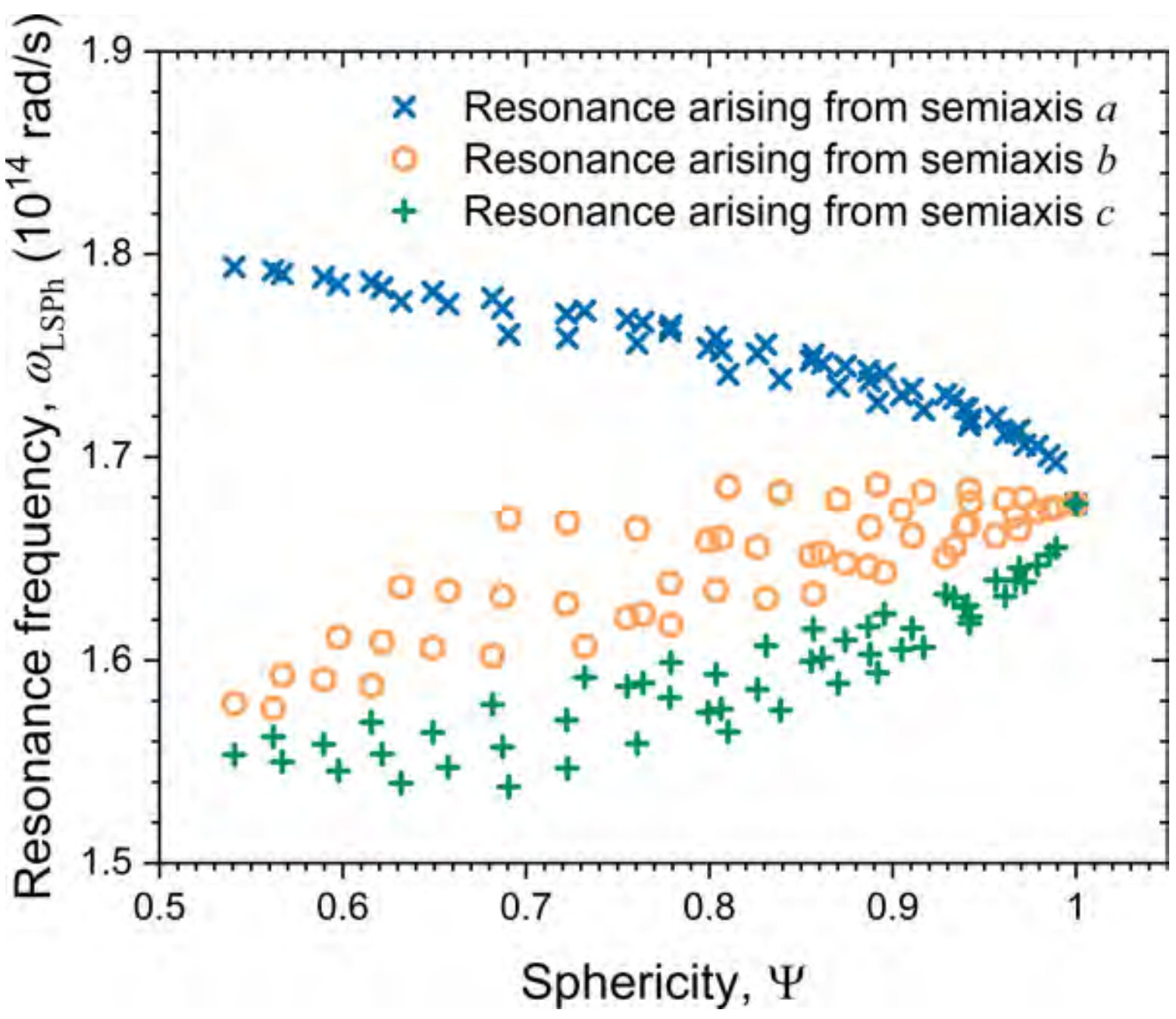

**Figure S3.** Resonance frequencies of LSPhs from a sampling of 55 ellipsoidal dipoles of variable sphericity. Ellipsoidal dipoles are made of SiC and embedded in a background reference medium with dielectric function $\varepsilon_{\mathrm{ref}} = 3$.

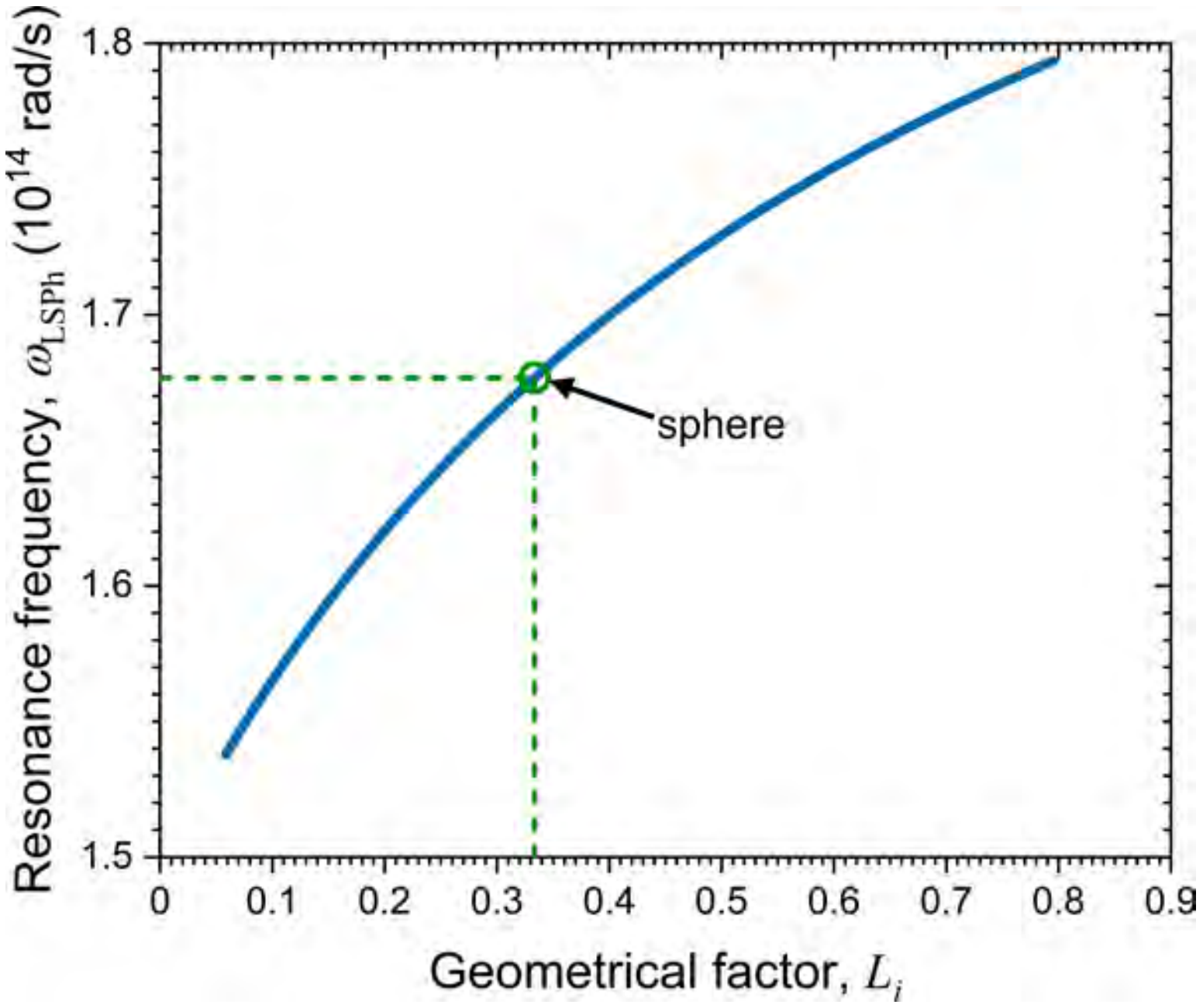

**Figure S4.** Resonance frequencies of LSPhs that correspond with the geometrical parameters $L_i$. Ellipsoidal dipoles are made of SiC and embedded in a background reference medium with dielectric function $\varepsilon_{\mathrm{ref}} = 3$. The geometrical parameter and corresponding resonance frequency for a sphere is depicted by the open green circle.

## S4. SPECTRAL ENERGY DENSITY ABOVE SiC METASURFACES WITH CONSTITUENT PARTICLES OF VARIABLE SPHRICITY AND $d = 6L_{\text{ch}}$ INTERPARTICLE SPACING

The spectral energy density above metasurfaces composed of particles of variable sphericity and structured with interparticle spacing $d = 6L_{\text{ch}}$ is presented in Fig. S5.

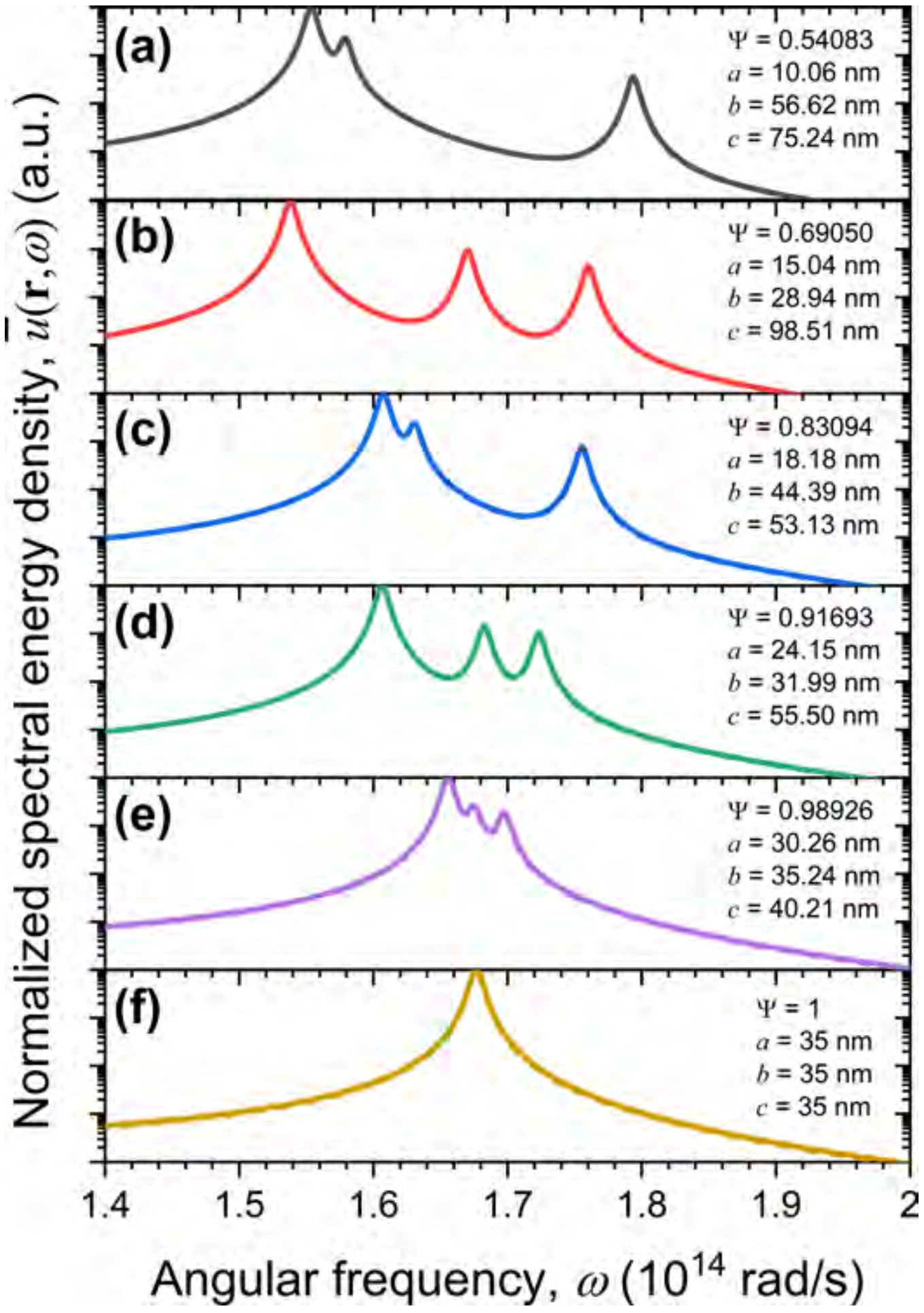

**Figure S5.** Normalized spectral energy density above metasurfaces with constituent particles of variable sphericity. Normalization is implemented as $\bar{u}(\mathbf{r},\omega) = u(\mathbf{r},\omega)/\max[u(\mathbf{r},\omega)]$ for each system. Each metasurface is composed of a 25-by-25 array of ellipsoidal SiC particles at temperature $T = 300$ K and embedded in a nonabsorbing background medium of $\varepsilon_{\text{ref}} = 3$. Solid lines represent metasurfaces and dashed lines represent a single particle. Here, the solid and dashed lines overlap. The sphericity of each particle is varied from $\Psi = 0.5408$ to $\Psi = 1$. All particles are unrotated. Interparticle spacing is $d = 6L_{\text{ch}}$. The observation point is located at a vertical distance of $d_{\text{obs}} = 2L_{\text{ch}}$ above the center of mass of the central particle in the array.

## S5. VISUALIZATION OF DISCRETE SYSTEM GREEN'S FUNCTION AT THE LOCALIZED SURFACE PHONON RESONANCE OF METASURFACES COMPOSED OF SPHERICAL PARTICLES

Fig. S6 visualizes the normalized absolute value of the discrete system Green's function $\bar{\bar{\mathbf{G}}}(\mathbf{r}_i, \mathbf{r}_j, \omega)$ components at the localized surface phonon (LSPh) resonance frequency $\omega_{\mathrm{LSPh}}$ of metasurfaces composed of spherical particles. Normalization is implemented as $\left|\mathrm{Re}\left[\bar{\bar{\mathbf{G}}}(\mathbf{r}_i, \mathbf{r}_j, \omega)\right]\right|/\max\left\{\left|\mathrm{Re}\left[\bar{\bar{\mathbf{G}}}(\mathbf{r}_i, \mathbf{r}_j, \omega)\right]\right|\right\}$ and $\left|\mathrm{Im}\left[\bar{\bar{\mathbf{G}}}(\mathbf{r}_i, \mathbf{r}_j, \omega)\right]\right|/\max\left\{\left|\mathrm{Im}\left[\bar{\bar{\mathbf{G}}}(\mathbf{r}_i, \mathbf{r}_j, \omega)\right]\right|\right\}$ for, respectively, the real and imaginary components of the discrete system Green's function. The discrete system Green's function for a metasurface constructed with interparticle spacing $d = 3L_{\mathrm{ch}}$ (Fig. S6(a)–(b)) is compared with that of a metasurface constructed with interparticle spacing $d = 6L_{\mathrm{ch}}$ (Fig. S6(c)–(d)).

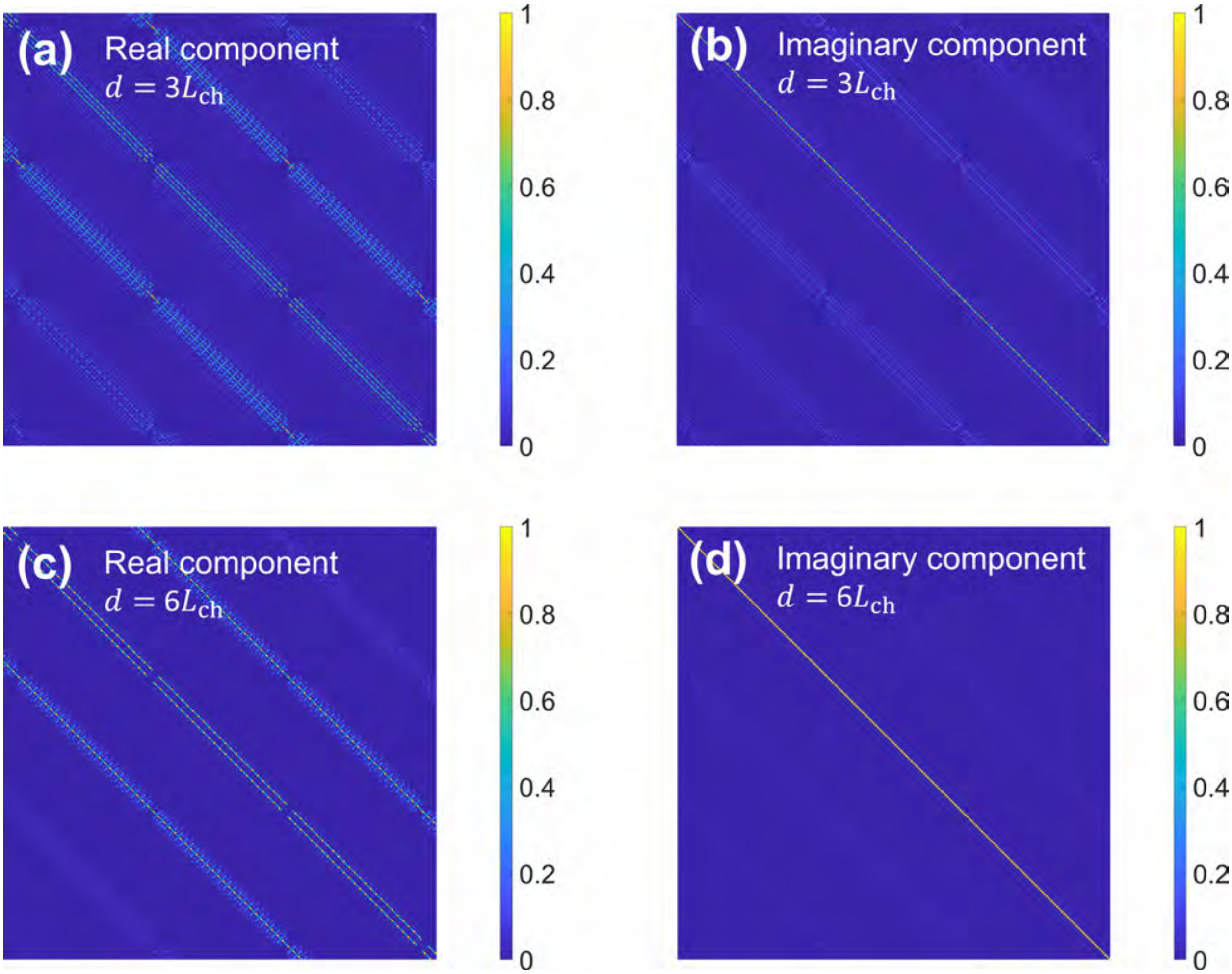

**Figure S6.** (a) Real and (b) imaginary components of the normalized discrete system Green's function at $\omega_{\mathrm{LSPh}} = 1.677 \times 10^{14}$ rad/s for a metasurface composed of a 25-by-25 array of spherical dipoles of radius $R = 35$ nm and with interparticle spacing $d = 3L_{\mathrm{ch}}$. (c) Real and (d) imaginary components of the normalized discrete system Green's function at $\omega_{\mathrm{LSPh}} = 1.677 \times 10^{14}$ rad/s for the same metasurface but with interparticle spacing $d = 6L_{\mathrm{ch}}$. The vertical axes represent index $i$ and the horizontal axes represent index $j$ in the discrete system Green's function $\overline{\overline{\mathbf{G}}}(\mathbf{r}_i, \mathbf{r}_j, \omega_{\mathrm{LSPh}})$. For visualization purposes, only the central 241-by-241 array of the complete 1875-by-1875 system Green's function matrices are modeled. Since there are $N = 25^2 = 625$ particles in the metasurface, the discrete system Green's function matrix has $3N \times 3N = 1875 \times 1875$ total terms, where the factor of three comes from the three Cartesian coordinates, $x$ , $y$, and $z$.

## S6. SPECTRAL ENERGY DENSITY ABOVE SiC METASURFACE WITH CONSTITUENT PARTICLES OF VARIABLE ORIENTATION AND $d = 6L_{\text{ch}}$ INTERPARTICLE SPACING

The spectral energy density above metasurfaces composed of particles of variable rotation angle and structured with interparticle spacing $d = 6L_{\text{ch}}$ is presented in Fig. S7.

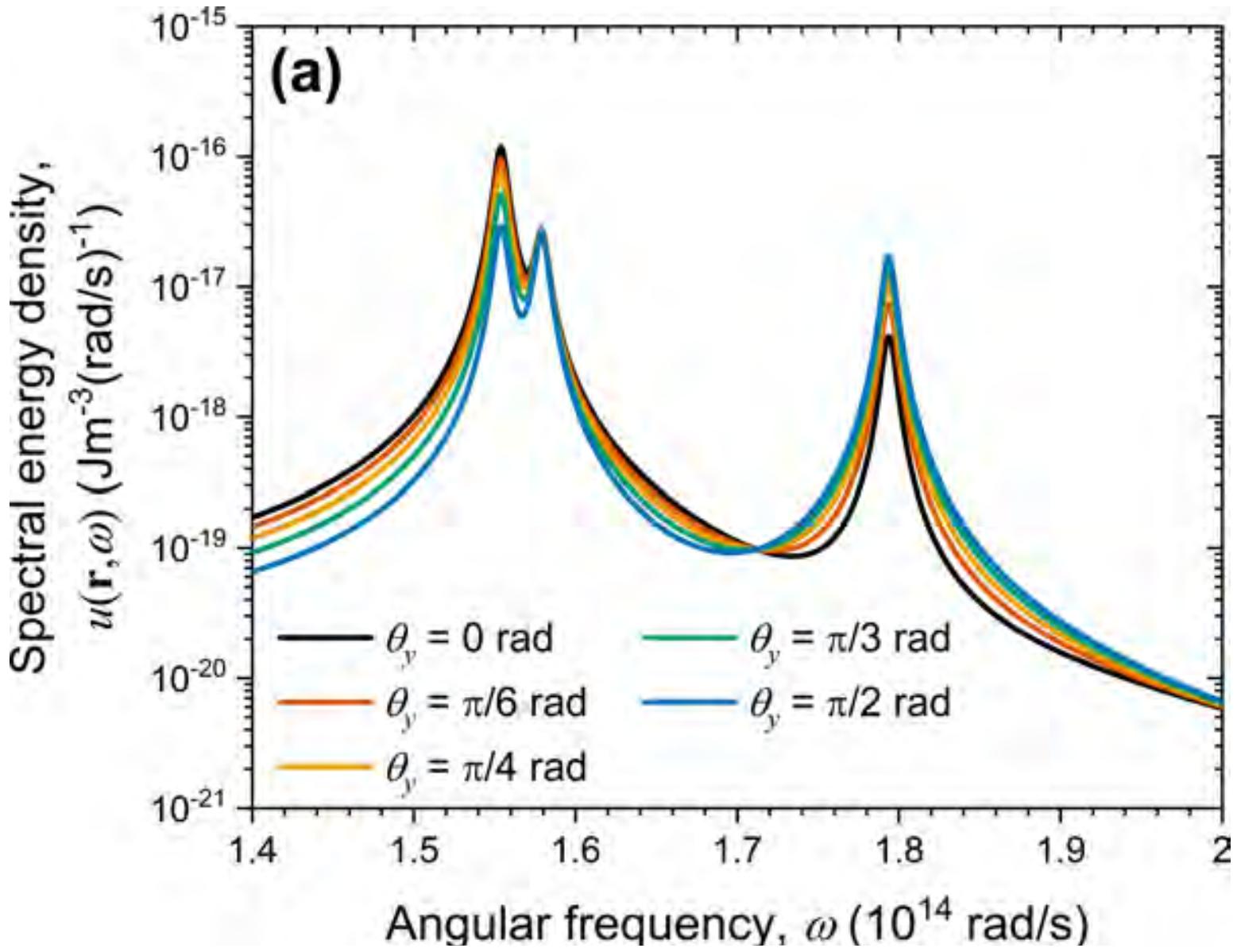

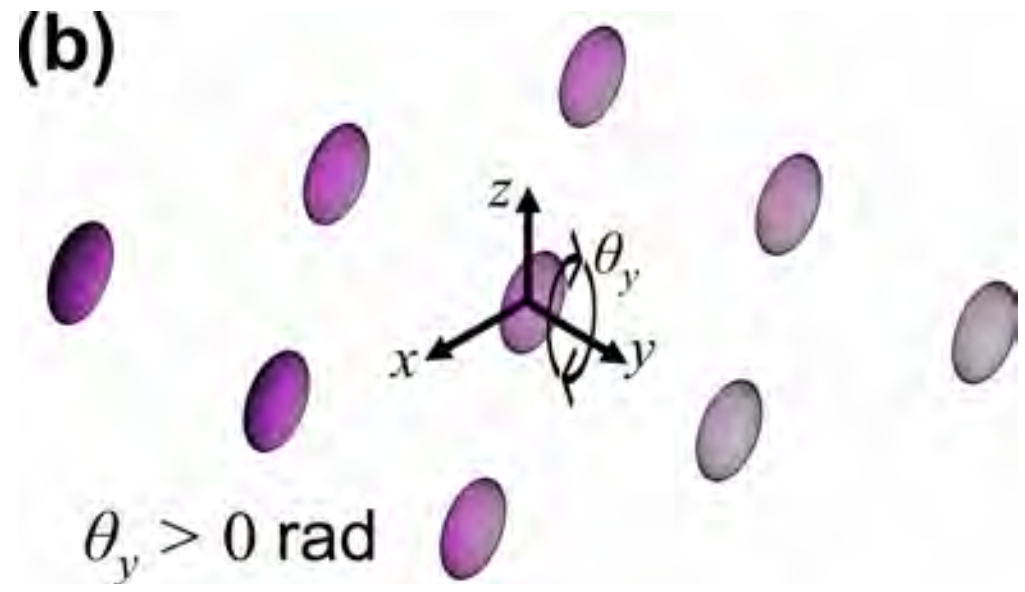

**Figure S7.** (a) Spectral energy density above metasurfaces with constituent particles of variable rotation angle. Each metasurface is composed of a 25-by-25 array of ellipsoidal SiC particles at temperature $T = 300$ K and embedded in a nonabsorbing background medium of $\varepsilon_{\text{ref}} = 3$. (b) Particles are rotated by the same angle $\theta_y$ in their local coordinate system and are of the same dimensions ($a = 10.06$ nm, $b = 56.62$ nm, $c = 75.24$ nm, $\Psi = 0.5408$). Interparticle spacing is $d = 6L_{\text{ch}}$. The observation point is located at a vertical distance of $d_{\text{obs}} = 2L_{\text{ch}}$ above the center of mass of the central particle in the array.

## S7. TOTAL ENERGY DENSITY ABOVE SiC METASURFACES WITH PARTICLES OF VARIABLE ORIENTATION

The total (i.e., spectrally integrated) energy density for the metasurface with constituent ellipsoidal particles of variable rotation angle and structured with interparticle spacing $d = 3L_{\mathrm{ch}}$ is presented in Fig. S8.

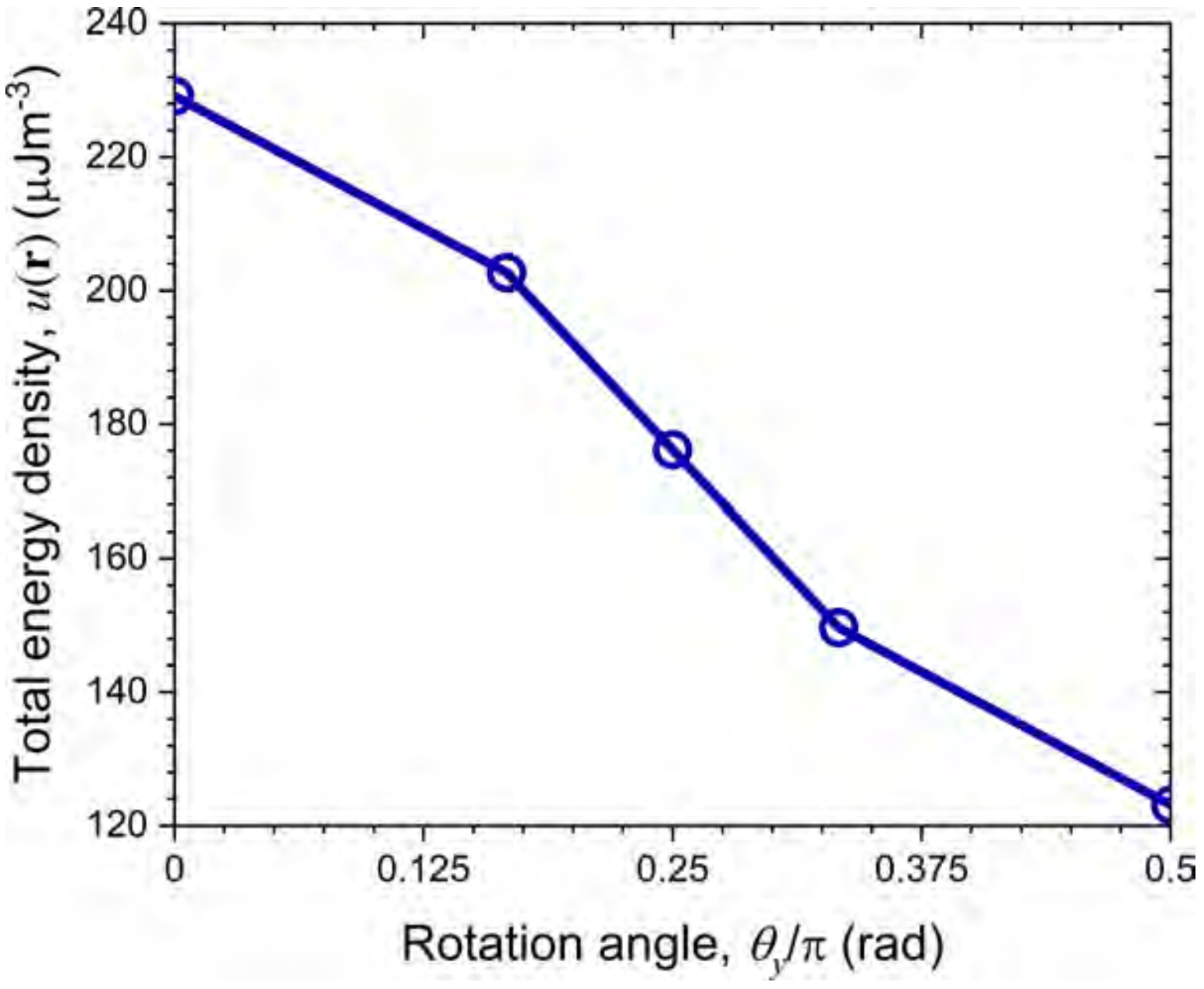

**Figure S8.** Total energy density above metasurfaces with constituent particles of variable rotation angle. Each metasurface is composed of a 25-by-25 array of ellipsoidal SiC particles at temperature $T = 300$ K and embedded in a nonabsorbing background medium of $\varepsilon_{\mathrm{ref}} = 3$. All particles are rotated by the same angle $\theta_y$ in their local coordinate system and are of the same dimensions ($a = 10.06$ nm, $b = 56.62$ nm, $c = 75.24$ nm, $\Psi = 0.5408$). Interparticle spacing is $d = 3L_{\mathrm{ch}}$. The observation point is located at a vertical distance of $d_{\mathrm{obs}} = 2L_{\mathrm{ch}}$ above the center of mass of the central particle in the array.

## S8. SPECTRAL ENERGY DENSITY ABOVE SiC METASURFACES WITH RANDOMIZED CONSTITUENT PARTICLES AND $\boldsymbol{d = 6L_{\mathrm{ch}}}$ INTERPARTICLE SPACING

The spectral energy density above metasurfaces constructed of ellipsoidal particles with random dimensions and random orientation is presented in Fig. S9. All metasurfaces are structured with interparticle spacing $d = 6L_{\mathrm{ch}}$.

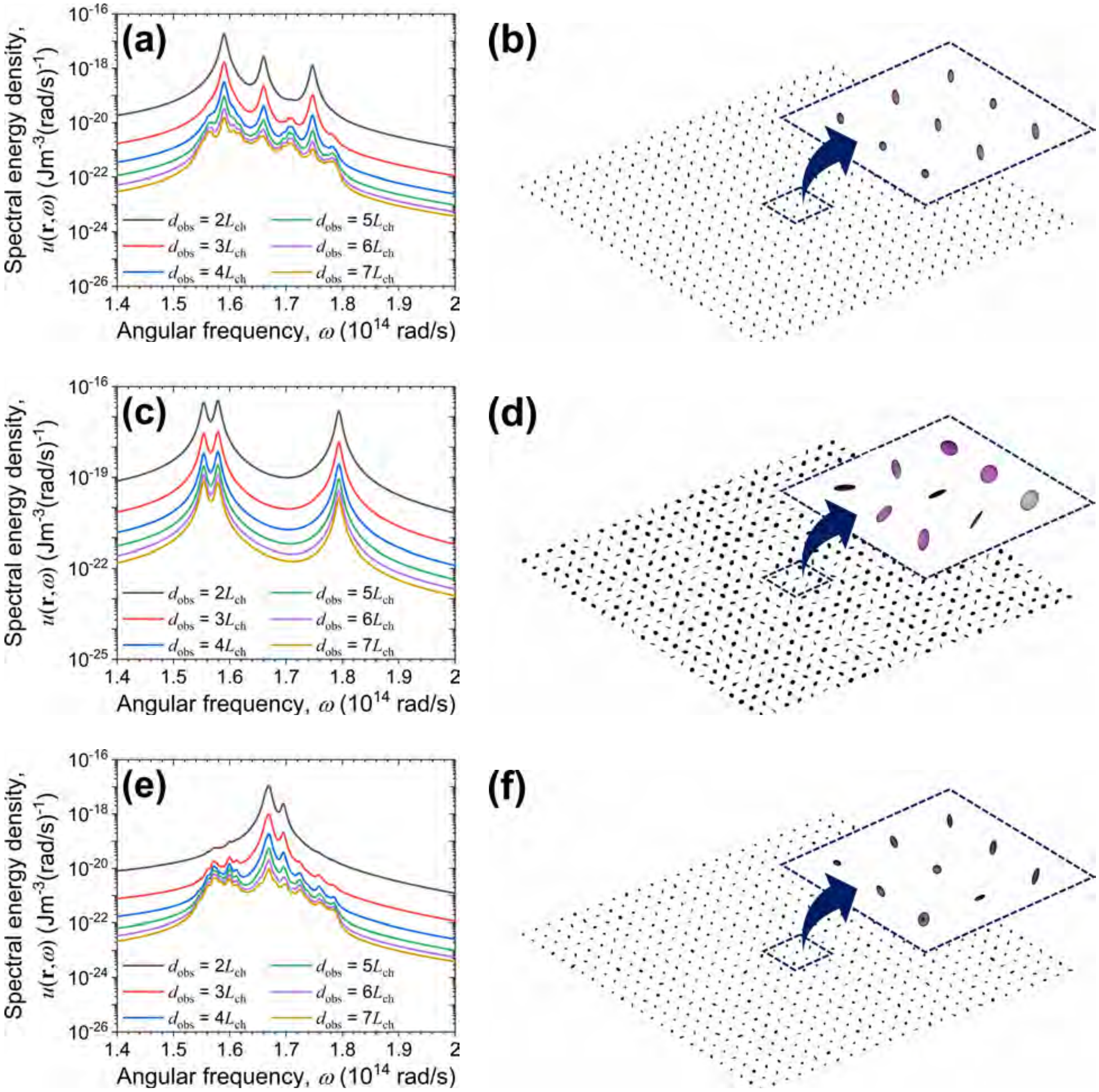

**Figure S9.** Spectral energy density above metasurfaces with randomized constituent particles. Each metasurface is composed of a 25-by-25 array of ellipsoidal SiC particles at temperature $T$ = 300 K and embedded in a nonabsorbing background medium of $\varepsilon_{\mathrm{ref}} = 3$. (a)–(b) Particles are unrotated and of random dimensions; (c)–(d) particles are all the same dimensions ($a$ = 10.06 nm, $b$ = 56.62 nm, $c = 75.24$ nm, $\Psi = 0.5408$) and of random orientation; and (e)–(f) particles are of random dimensions and random orientation. Interparticle spacing is $d = 6L_{\mathrm{ch}}$. Spectral energy density is measured at six different observation points located at vertical distances of $d_{\mathrm{obs}} = 2L_{\mathrm{ch}}$, $3L_{\mathrm{ch}}$, $4L_{\mathrm{ch}}$, $5L_{\mathrm{ch}}$, $6L_{\mathrm{ch}}$, $7L_{\mathrm{ch}}$ above the center of mass of the central particle in the array.